\documentclass[11pt,a4paper]{article}
\pdfoutput=1
\usepackage{jheppub}

\usepackage{amsmath}
\usepackage{amssymb}
\usepackage{epsfig,multicol,bbm}
\usepackage{graphicx}
\usepackage{tabu}
\usepackage{multirow}
\usepackage{bbold}
\usepackage[mathscr]{euscript}
\usepackage{makecell}
\usepackage[normalem]{ulem}

\newcommand{\be}[0]{\begin{equation}}
\newcommand{\ee}[0]{\end{equation}}

\newcommand{\eq}[1]{Eq.~\eqref{eq:#1}}

\newcommand{\bea}{\begin{eqnarray}}
\newcommand{\eea}{\end{eqnarray}}

\newcommand{\nn}{\nonumber}

\DeclareMathOperator{\Tr}{Tr}

\newcommand{\as}{\alpha_s}

\newcommand{\unmeas}{ {\rm unmeas}}

\newcommand{\lb}{\Big{\lbrack}}
\newcommand{\rb}{\Big{\rbrack}}
\newcommand{\lp}{\Big{(}}
\newcommand{\rp}{\Big{)}}
\newcommand{\lbc}{\Big{\lbrace}}
\newcommand{\rbc}{\Big{\rbrace}}
\newcommand{\veb}[1]{#1_{T}^{\mathrm{cut}}}
\newcommand{\vebp}[1]{#1_{\perp}^{\mathrm{cut}}}
\newcommand{\vebc}[1]{#1^{\mathrm{cut}}}
\newcommand{\vebv}[1]{\boldsymbol{\mathrm{#1}}_{\perp}}
\newcommand{\prp}[1]{#1_{\perp}}
\newcommand{\tra}[1]{#1_{T}}
\newcommand{\bmat}[1]{\boldsymbol{\mathrm{#1}}}
\newcommand{\overbar}[1]{\mkern 1.5mu\overline{\mkern-1.5mu#1\mkern-0.3mu}\mkern 1.5mu}
\newcommand{\bigv}{\Big{\vert}}
\newcommand{\lngl}{\Big{\langle}}
\newcommand{\rngl}{\Big{\rangle}}

\title{Transverse Vetoes with Rapidity Cutoff in SCET}

\author[a]{Andrew Hornig, }
\author[a,b]{Daekyoung Kang, }
\author[a,c]{Yiannis Makris, }
\author[c]{and Thomas Mehen}

\affiliation[a]{Theoretical Division T-2, Los Alamos National Laboratory, Los Alamos, NM, 87545}
\affiliation[b]{Key Laboratory of Nuclear Physics and Ion-beam Application (MOE) and Institute of Modern Physics, Fudan University, Shanghai, China 200433}
\affiliation[c]{Department of Physics, Duke University, Durham, NC 27708}

\emailAdd{andrew.hornig@gmail.com}
\emailAdd{dkang@fudan.edu.cn}
\emailAdd{yiannis@lanl.gov}
\emailAdd{mehen@phy.duke.edu}

\abstract{We consider di-jet production in hadron collisions where a transverse veto is imposed on radiation for (pseudo-)rapidities in the central region only, where this central region is defined with rapidity cutoff. For the case where the transverse measurement (e.g., transverse energy or min $p_T$ for jet veto) is parametrically larger relative to the typical transverse momentum beyond the cutoff, the cross section is insensitive to the cutoff parameter and is factorized in terms of collinear and soft degrees of freedom. The  virtuality for these degrees of freedom is set by the transverse measurement, as in typical transverse-momentum dependent observables such as Drell-Yan, Higgs production, and the event shape broadening. This paper focuses on the other region, where the typical transverse momentum below and beyond the cutoff is of similar size. In this region the rapidity cutoff further resolves soft radiation into (u)soft and soft-collinear radiation with different rapidities but identical virtuality. This gives rise to rapidity logarithms of the rapidity cutoff parameter which we resum using renormalization group methods. We factorize the cross section in this region in terms of soft and collinear functions in the framework of soft-collinear effective theory,  then further refactorize  the soft function as a convolution of the (u)soft and soft-collinear functions. All these functions are calculated at one-loop order. As an example, we calculate a differential cross section for a specific partonic channel, $q q' \to q q'$, for the jet shape angularities and show that the refactorization allows us to resum the rapidity logarithms and significantly reduce theoretical uncertainties in the jet shape spectrum.}

\keywords{Jets, Factorization, Resummation, Effective Field Theory}
\preprint{LA-UR-17-27349}

\begin{document}
\maketitle
\noindent


\section{Introduction}
\label{sec:intro}

 In the recent years, jet substructure has been of great interest to the particle physics community since it can be used to discriminate between jets of different origins, e.g., quark and gluon jets or jets from hadronic decays of boosted heavy mesons or Higgs and $Z$ bosons ~\cite{Chien:2012ur,Dasgupta:2013ihk, Jouttenus:2013hs,Larkoski:2014wba, Larkoski:2014uqa, Larkoski:2013paa, Larkoski:2013eya, Dasgupta:2012hg,Dasgupta:2013via,Becher:2015gsa,Chien:2014nsa,Larkoski:2015kga}. This is essential for expanding  our understanding of quantum chromodynamics (QCD) as well as testing the standard model (SM) or searching for beyond SM physics.

 In  experimental studies of  exclusive $N$-jet production, it is common to impose a veto on the out-of-jet radiation  in order to control soft emissions. Additionally, due to detector limitations these vetoes are imposed within a specific (pseudo-)rapidity region and  the veto is not imposed outside this region. These constraints can induce large logarithms of the cutoff parameter, $e^{-\vebc{\eta}}$, and ratios of the veto parameter, $\veb{p}$, to other scales in the problem (e.g., the hard scale $\mu_{H} \sim \sqrt{-t}$, where $t$ is the usual Mandelstam variable). These logarithmic enhancements  could potentially ruin the effectiveness of the ordinary perturbative expansion. In this work we propose a factorization theorem  for resumming logarithms of $e^{-\vebc{\eta}}$ and $\veb{p}/\mu$ within the framework of soft-collinear effective theory (SCET)~\cite{Bauer:2000ew,Bauer:2000yr,Bauer:2001ct,Bauer:2001yt}, where $\mu$ is the factorization scale. SCET was extensively used in the past decade for factorization of observables with sensitivity to soft and collinear radiation, such as jet substructure measurements in hadronic colliders. Some other interesting  applications of SCET include  cross sections for event shapes in the collinear limit, jet production rates~\cite{Bauer:2002ie, Bauer:2008dt}, and identified hadrons within jets~\cite{Jain:2011xz,Procura:2009vm,Liu:2010ng,Baumgart:2014upa,Bain:2016clc,Dai:2016hzf, Kang:2016ehg, Ritzmann:2014mka}.

 In this paper, we study rapidity cutoff resummation and  develop the necessary ingredients for di-jet cross sections with transverse energy, $\tra{E}$, and jet-veto measurements in hadronic collisions. Our results can easily be extended to zero and one-jet cross sections as well. The transverse energy is defined as the sum of the scalar transverse momentum of all the particles that do not belong to a jet and have rapidity, $\eta$, in the range $ \vert \eta \vert < \vebc{\eta}$, 
\begin{equation}
  \tra{E} = \sum_{i \notin \text{jet}} \vert \tra{\bmat{p}^i} \vert \Theta( \vebc{\eta} - \vert \eta_{i} \vert),
\end{equation}
where the rapidity is measured with respect to the beam axis. The veto is implemented by imposing the constraint $\tra{E} < \veb{p}$.  Due to the nature of the observable one expects that such a measurement is sensitive to the underlying event (UE). Measurements of the UE activity have been performed by ATLAS and CMS in inclusive charged particle production \cite{Aaboud:2017fwp}, Drell-Yan \cite{Aad:2016ria,Chatrchyan:2012tb,CMS:2016etb}, and exclusive dijet events \cite{Aad:2014hia}.

The effect of UE in transverse energy resummed distributions was studied in Refs.~\cite{Papaefstathiou:2010bw,Grazzini:2014uha} for $\eta^{\text{cut}} = 4.5$ in the case of Higgs and vector-boson production using monte-carlo simulation.  In contrast to the work in this paper, in Refs.~\cite{Papaefstathiou:2010bw,Grazzini:2014uha} the rapidity cutoff was introduced only during the simulation and not in the resummed distribution. However, as will be discussed below, for the large values of rapidity cutoff ($\eta^{\text{cut}} \gtrsim 4.5$) the effect of the cutoff on the resummed distribution is expected to be small. In this work we ignore effects of multiparton interactions and focus on contributions from initial and final state radiation. In principle, effects from multiparton interactions could be included later on top of our analysis as factorization breaking corrections but this is beyond the scope of this work. 

Similarly, the jet-veto measurement imposes $\vert \tra{\bmat{p}}^{i}(R^{\text{veto}}) \vert <\veb{p}$, where $\tra{\bmat{p}}^{i}$ is the transverse momenta of the $i$-th jet reconstructed by a jet algorithm and $R^{\text{veto}}$ is the jet cone size parameter used during the vetoing process which could be different from the hard jet size, $R$. Though the jet-veto measurements are less sensitive to UE, they suffer from logarithmic enhancements of $R^{\text{veto}}$. Such logarithms are known as ``clustering logarithms'' and appear in next-to-next-leading order (NNLO) calculations ~\cite{Tackmann:2012bt,Banfi:2012yh,Stewart:2013faa}, and could make an important contribution to the cross section. At present there is no known  method for resummation of these logarithms but they could be included order-by-order in perturbation theory.   

As a preliminary exercise we study di-jet production under the rapidity constraints in an electron-positron annihilation process. Specifically we study the effects of the rapidity constraints in the small transverse energy regime $\Lambda_{\text{QCD}} \ll \prp{E} \lesssim \omega r \ll \omega$, where $\omega = \sqrt{s}$ is the center of mass energy and $r=e^{-\vebc{\eta}}$ is the rapidity cut. For this simple example, the transverse energy $\prp{E}$ as well as the rapidity is measured with respect to the thrust axis and therefore we use different notation ($\prp{E}$ instead of $\tra{E}$) to avoid confusion. The schematic form of the factorization of the cross section within SCET is,
\begin{align}
  \frac{d\sigma}{d\prp{E}} &\sim H\times S \otimes \mathcal{J}_q \otimes \mathcal{J}_{\bar{q}} , & S(\prp{E}) &= S_{s}\otimes S_{n} \otimes S_{\bar{n}},
\end{align}
where $\otimes$ denotes convolution over $\prp{E}$. The hard function, $H$ describes the hard process: $e^+ e^- \to q\bar{q}$, and the soft function, $S$, describes the soft radiation and cross talk between collinear sectors. The collinear radiation along the thrust axis is described by the functions $\mathcal{J}_{q}$ which can be written in terms of the ``unmeasured'' jet function\footnote{We use the terminology of Ref.~\cite{Ellis:2010rwa} and we refer to jets for which no substructure observable is measured as unmeasured jets }, $J_{i}$, introduced in Refs.~\cite{Ellis:2010rwa}, and contributions from out-of-jet radiation, which we denote as $\Delta J_{i}$. For small values of the transverse momentum, $\prp{E} \ll \omega r $, the collinear radiation which is emitted within the cone has parametrically large transverse momenta, compared to $\prp{E}$, and does not contribute to the measurement. In this case the function $\mathcal{J}_q$ reduces to the standard unmeasured jet function. The corrections from the out-of-jet radiations are necessary to describe the process for moderate values of $\prp{E}\lesssim \omega r$. The collinear-soft function, $S_n$, describes the collinear-soft modes which are collinear in the $n$-direction and therefore can resolve the jet-cone boundary. The global-soft function, $S_{s}$, describes the standard u-soft modes of SCET$_{\text{I}}$ which cannot resolve the small jet radius and therefore for the calculation of $S_{s}$ no rapidity constraints are imposed. In this $e^+e^-$ example there is no UE and therefore the factorization is accurate up to higher orders in the effective field theory power counting parameter $\lambda \sim \prp{E}/\omega$. This allows us to directly compare our results with simulation data. For our analysis we use MadGraph~\cite{Alwall:2014hca} $+$ \textsc{Pythia} 8~\cite{Sjostrand:2006za,Sjostrand:2007gs}.  Our calculations are in very good agreement with Monte Carlo for most values of $E_\perp$.

The refactorization of the soft function into global-soft and collinear-soft terms introduces rapidity divergences which we regulate using the rapidity regulator of Refs.~\cite{Chiu:2012ir,Chiu:2011qc}. The rapidity scale dependence allow us to derive rapidity renormalization group (RRG) equations which we solve to resum global logarithms of $r$  up to next-to-leading logarithmic (NLL) accuracy. This process closely follows the analysis in Refs.~\cite{Becher:2015hka, Chien:2015cka} where resummation of jet size parameter is performed in the context of electron-positron annihilation and in Refs.~\cite{Hornig:2016ahz,Bertolini:2017efs} for proton-proton collisions. Non-global logarithms (NGLs)~\cite{Dasgupta:2001sh,Banfi:2002hw,Appleby:2003ai,Dokshitzer:2003uw,Khelifa-Kerfa:2015mma,Hornig:2011tg,Hornig:2011iu, Kelley:2011aa} of $r$ appear at NNLO calculations. Their resummation is particularly challenging since they do not have the same pattern at each order in perturbative expansion. 
NGLs can be included order by order in $\as$ when their contribution is not large. Otherwise, resummation strategies developed in perturbative QCD \cite{Dasgupta:2001sh,Banfi:2002hw,Weigert:2003mm,Avsar:2009yb,Hatta:2013iba,Marchesini:2003nh,Forshaw:2009fz, DuranDelgado:2011tp}  or  recent approaches  in the framework of SCET \cite{Larkoski:2015zka,Becher:2015hka,Neill:2015nya,Becher:2016mmh,Becher:2017nof} should be adopted.

Other logarithms of ratios of widely separated scales  also appear in the factorized cross section (e.g. ratios of $\mu_H = \omega$, $\mu_J = \omega r$, and  $\mu_{ss} = \prp{E}$) are resummed by using the standard RG evolution within the effective theory. We summarize the RG evolution properties for all relevant terms in the Sections~\ref{sec:2.2}, \ref{sec:3.2}, and  Appendix~\ref{app:A.3}. 

In hadronic collisions the beam direction plays the role that thrust axis plays in electron-positron collisions and the collinear radiation along the $n_{B}$-direction is described by the beam functions~\cite{Stewart:2009yx}. The corrections to the beam function from out-of-beam radiation will contribute to the transverse energy (or jet-veto) measurement in a similar way as corrections  to the jet function in electron-positron annihilation.  For jet production, in addition to the $n_B$-collinear-soft and $n_B$-collinear  modes, we also have corresponding modes along each jet direction.  The contribution from $n_J$-collinear-soft modes is considered through further refactorization of the soft function including the $n_J$-collinear-soft function, $S_{n,J}$. The factorization theorem for $N$-jet production in hadronic collisions is,
\begin{equation}\label{eq:sigmanjet}
  \frac{d\sigma}{d\tra{E}} \sim  \Tr \lb \bmat{H}_{ab \to 1,2...,N }\; \bmat{S}^{ab \to 1,2...,N}_{\text{unmeas}} \rb \otimes \mathcal{B}_{a/P} \otimes \mathcal{B}_{b/P} \times
  \lp \prod_{i=1}^{N}  J_{i} \rp,
\end{equation}
where
\begin{equation}
  \bmat{S}^{ab \to 1,2...,N}_{\text{unmeas}} = \bmat{S}_s^{ab \to 1,2\ldots, N} \otimes S_{n,B}^{(a)}\otimes S_{n,\overbar{B}}^{(b)} \otimes  S_{n,J}^{(1)} \cdots \otimes  S_{n,J}^{(N)},
\end{equation}
and $\bmat{S}_s$ is the global $N$-jet soft function, $S_{n,B}^{(a)}$, and $S_{n,J}^{(i)}$ are the collinear soft functions along the beam and jet directions respectively. The superscripts $(a)$ and $(i)$ denote the partons associated with these functions and it should be noted that $S_{n,B}^{(a)}$ and $S_{n,J}^{(i)}$ are different functions even for the same parton $a=i$ because the veto for both functions is always applied respect to the beam direction, not individual jet and beam directions. The parton dependence of the soft functions  will be suppressed for the rest of the text for simplicity of notation. The contributions from the $n_J$-collinear modes to the transverse energy are suppressed and therefore the jet functions $J_{i}$ do not participate in the convolutions over $\tra{E}$, while the beam functions $\mathcal{B}_{i/P}$ do through $\tra{E}/\omega r$ terms in the power corrections. This is discussed in Appendix~\ref{app:C}. It should be noted that in this paper we focus in the region $\tra{E} \sim \omega r$ where these corrections are important and NGLs of the form $\ln (\tra{E}/\omega r) \sim 1$ and therefore resummation is not needed.

The factorization formula in \eq{sigmanjet} assumes the jets are in the central rapidity region,
hence the factorization is invalid for jets in the large rapidity region, for which pure $t$-channel forward scattering dominates. This process was extensively studied in the framework of High Energy Jets (HEJ) \cite{Alioli:2012tp, Andersen:2011hs, Andersen:2009nu} developed to resum logarithms of the rapidity difference between jets and in the context of factorization violation by Glauber-gluon exchange  \cite{Rothstein:2016bsq}. 
The formula  in \eq{sigmanjet} can be used to calculate exclusive N-jet production in the absence of UE and to understand its effect by comparing to ATLAS and CMS measurements \cite{Aad:2016ria,Chatrchyan:2012tb,CMS:2016etb,Aad:2014hia}.

We also consider jet substructure measurements for some of the jets. In this case the factorization is obtained with the replacement~\cite{Hornig:2016ahz,Ellis:2010rwa},
\begin{equation}
   \frac{d\sigma}{d\tra{E} d\tau} \sim  \frac{d\sigma}{d\tra{E}} \lp J_{i} \to S_{\text{meas}} \otimes_{\tau} J_{i}(\tau)  \rp,
\end{equation}
where $S_{\text{meas}}(\tau)$ is the contribution of the collinear-soft modes to the measurements within the jet cone, $ J_{i}(\tau)$ is the measured jet function, and $\otimes_{\tau}$ denotes a convolution over the jet substructure observable, $\tau$. The unmeasured soft function, $\bmat{S}_{\text{unmeas}}$, is universal, meaning it does not depend on the jet substructure observable and therefore our calculations hold for jet substructure studies as well.


In Section~\ref{sec:ee} we motivate our analysis using the simple example of electron-positron annihilation and we show two distinct factorization theorems are required to explain the simulation data in the region $\prp{E} \ll \omega$. One factorization theorem involves the inclusive soft and jet functions, does not depend on the rapidity cutoff, and describes the region $ \omega r \ll \prp{E} \ll \omega$. On the other hand for small values of transverse energy, $\prp{E} \ll \omega r \ll \omega$, we find that a factorization is needed which is sensitive to the rapidity cutoff. In this section we also calculate the correction to the unmeasured jet function from out of jet radiation  and demonstrate that including such corrections greatly improves the agreement with the simulations for  $\prp{E} \lesssim \omega r$. In Section~\ref{sec:pp} we extend the formalism to hadronic collisions and we give all perturbative matching coefficients for the unmeasured quark beam functions and the corrections from out-of-beam radiation. The  details of the calculation for the matching coefficients are given in Appendix~\ref{app:B}. In Section~\ref{sec:pp} we also construct the universal part of the di-jet soft function and we discuss the RG evolution of the soft function in rapidity and virtuality space for both transverse energy and jet-veto measurements. In Section~\ref{sec:app} we apply this formalism to the example of di-jet production with one unmeasured jet and  one for which the angularity (see Ref.~\cite{Hornig:2016ahz}) $\tau_{0}$ is measured. We impose a jet-veto measurement on the  soft out-of-jet radiation, and we  focus on small, $\tau_0$, region for which angularity is approximately proportional to the jet invariant mass,
\begin{equation}
\tau_0 = m_J^2/\tra{p}^2+ \mathcal{O}(\tau_0^2),
\end{equation}
where $\tra{p}$ is the jet transverse momentum measured from the beam axis. Even though the calculation of di-jet cross section involves summing over all possible partonic channels here we consider only the case $qq' \to qq'$. The complete calculation is beyond the scope of this paper.  We conclude in Section~\ref{sec:conclusion}.


\section{Two jets with an $E_\perp$ veto in electron-positron annihilation}
\label{sec:ee}

\begin{figure}[t!]
  \centerline{\includegraphics[width = \textwidth]{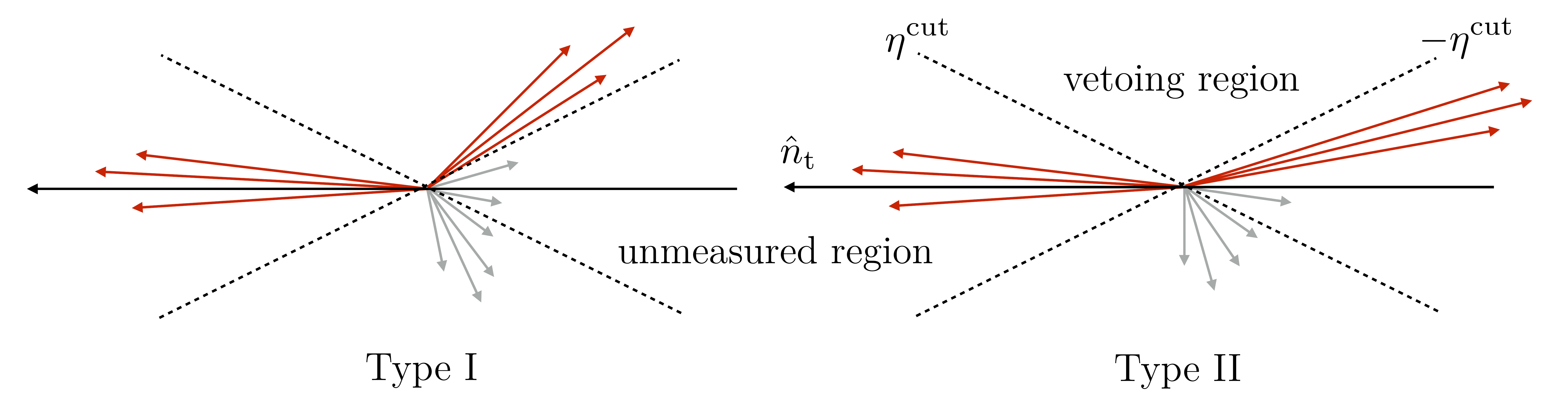}}
\caption{Topology of Type I and Type II events. The vetoing region $\vert \eta \vert < \vebc{\eta}$ is defined as the area in which we perform the measurement of the quantity which will be using to perform the transverse veto. Type I events contain at least one collinear sector in this region where Type II contain no such collinear sectors.}
\label{fig:fig1}
\end{figure}

To obtain a better understanding of the effect of a soft radiation veto with rapidity constraints in a setting simpler than hadron-hadron collisions, we study a similar observable  in electron-position annihilation. To define the veto we use the rapidity of final state particles measured with respect to the thrust axis. At leading order in the strong coupling the cross section is dominated by di-jet events where the two-jet axis is close  to the thrust axis. We can then categorize the events into two types. In the first category which we refer to as Type-I events, at least one of the two collinear sectors points inside the vetoing area as shown in Fig.~\ref{fig:fig1} (left). The second category which we call Type-II events both collinear sectors point outside the vetoing area as shown in Fig.~\ref{fig:fig1} (right).

\begin{figure}[t!]
  \centerline{\includegraphics[width = \textwidth]{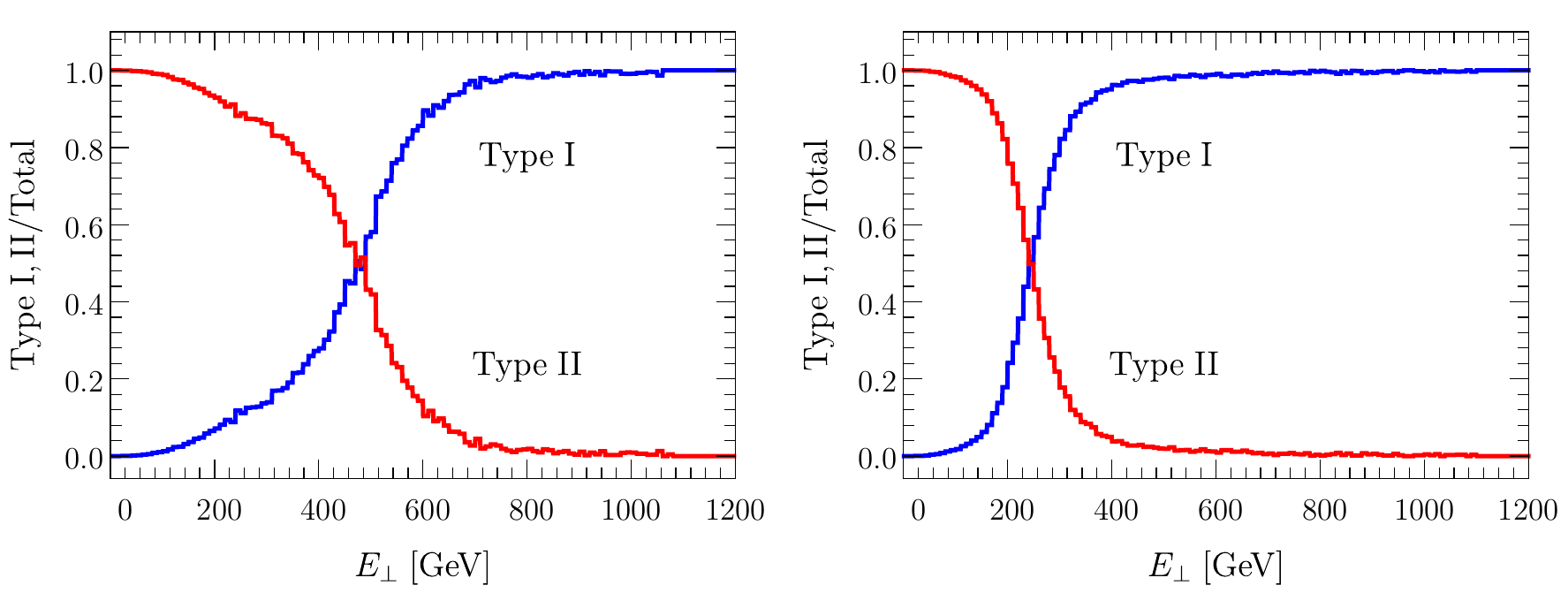}}
\caption{The ratio of the number of events of Type-I (Type-II) divided the total number of events for each bin of the transverse energy is plotted in blue (red) from \textsc{Pythia} simulations. We analyzed center of mass events for $\sqrt{s}=\omega=2$ TeV and for $\vebc{\eta} =1.5$ (left) where $\omega r \sim 450$ GeV and $\vebc{\eta} =2.5$ (right) where $\omega r \sim 164$ GeV. The collinear sectors are identified in $\textsc{Pythia}$ using anti-$k_T$ \textsc{FastJet} analysis for narrow jets ($R=0.05$). }
\label{fig:trans}
\end{figure}

We consider a veto using the measurement of transverse energy, $\prp{E}$, which is defined as the scalar sum of the transverse momentum of all particles in the pseudo-rapidity region $\vert \eta \vert < \vebc{\eta}$:
\begin{equation}
  \prp{E} = \sum_i \vert \vebv{p}^i \vert \Theta(\vebc{\eta} - \vert \eta_i \vert) ,
\end{equation}
where the sum extends over all particles in the event.  We are looking for the hierarchy between $E_{\perp}$, $\omega r $, and $\omega$  that will allow us to separate regions of the phase space where either Type-I  or Type-II events dominate the cross section, where $r$ is defined by $r = \exp (- \vebc{\eta})$~\footnote{The parameter $r$ is also related to the half opening angle, $\phi$, of the  cones: $r=\tan(\phi/2)$}.
 We study the fraction of Type-I and Type-II events as a function of transverse energy, $E_{\perp}$, for different values of $\eta_{\text{cut}}$ at $\omega = 2$ TeV using \textsc{Pythia} 8.  In our simulations we have turned off hadronization. The thrust axis is defined globally, and then the anti-$k_T$ algorithm is used to find jets with $R=0.05$. We require that the two most energetic jets carry 90\% of the total energy. If both of these jets are outside the veto the event is Type-II, otherwise it is Type-I. The results for $\eta_{\text{cut}}=1.5$ and $\eta_{\text{cut}}=2.5$  are presented in Fig.~\ref{fig:trans}.

From these plots we find that for $\omega r \ll E_{\perp}$, Type-I events dominate the cross section, where for  $E_{\perp} \ll \omega r $, Type-II events dominate. This can be understood from  basic kinematics. In each hemisphere the total transverse momentum is zero (from the definition of the thrust axis) thus the collinear radiation will recoil against the soft radiation which is emitted at larger angles. The transverse momentum of the collinear sector at the transition point from Type-I to Type-II is given by $E^{\text{coll.}}_{\perp} \sim \omega r $ therefore at this region of phase-space $E_{\perp} = E_{\perp}^{\text{coll.}}+E_{\perp}^{\text{soft}} \gtrsim 2 \omega r$. On the other hand, transitioning from Type-II to Type-I events, where $E_{\perp}=E_{\perp}^{\text{soft}}$~\footnote{ Note that $E_{\perp}$ is measured only in the vetoing area thus for Type-II events only the soft radiation contributes to the measurement. For Jet-veto type measurements (rather than $E_{\perp}$-veto) both transitions happen at the same point.}, the transition begins at $E_{\perp}\lesssim \omega r$.

\begin{figure}[t!]
  \centerline{\includegraphics[width = \textwidth]{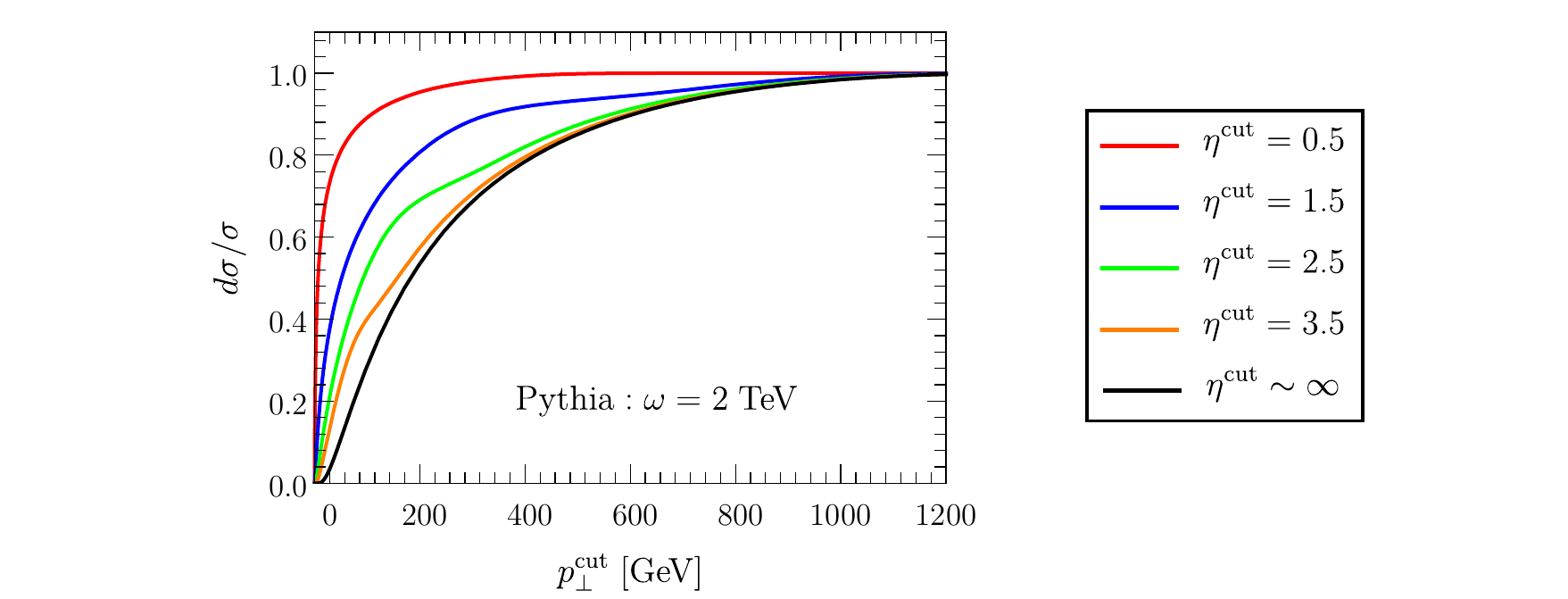}}
\caption{The integrated cross section as a function of the vetoing parameter $\veb{p}$ from \textsc{Pythia} simulation for various values of $\vebc{\eta}$: $\vebc{\eta}=0.5~\text{(red)}$ , $1.5~\text{(blue)}$, $2.5~\text{(green)}$, $3.5~\text{(orange)}$. We also give the results for no rapidity constraints, $\vebc{\eta} \to \infty$, (black). }
\label{fig:con}
\end{figure}

Therefore we identify two different regions and construct factorization theorems for each region within the framework of SCET. The two regions of phase-space we are considering are:
\begin{align}
  \mathrm{Region \; I:}\;\;\; &\omega r \ll \prp{E} \ll \omega \nn \\
  \mathrm{Region \; II:}\;\;\; &\prp{E} \ll \omega r  \ll \omega\; .
\end{align}
Region I which is dominated by Type-I events which have no sensitivity to the exact value of $r$ as long as it respects the hierarchy of scales that describes this region. The reason is that the modes sensitive to the size of $r$ are soft and collinear with typical transverse momentum $\sim \prp{E} r $ and therefore contribution of such particles to the measurement is parametrically small. This suggests that for Type-I events we can take $\eta_{\rm cut}\to \infty$ and obtain a good approximation to the cross section. To verify this, we define the integrated cross section
\begin{equation}
  \label{eq:cum}
  d\sigma(\vebc{\eta},\vebp{p}) \equiv \int_{0}^{\vebp{p}} d\prp{E} \frac{d\sigma}{d\prp{E}}(\vebc{\eta},\prp{E}).
\end{equation}
and calculate the ratio of this cross section to the total cross section as a function of $\vebp{p}$ for various values of $\vebc{\eta}$, as well as $\vebc{\eta}\to\infty$, in Fig.~\ref{fig:con} using \textsc{Pythia} simulations. We find that for sufficiently large values of $\vebp{p}$  the cross section $d\sigma(\vebc{\eta},\vebp{p})$ asymptotically approaches  $d\sigma(\vebc{\eta} \to \infty,\vebp{p})$. Note that the finite $\vebc{\eta}$ curves approach the $\vebc{\eta}\to\infty$ curve when $\vebp{p} \sim \omega r$.  

This approximation was discussed in Refs.~\cite{Tackmann:2012bt, Gangal:2016kuo} and has been used in subsequent studies~\cite{Stewart:2013faa,Kolodrubetz:2016dzb} of various jet observables at hadron colliders within the framework of SCET$_{\text{II}}$. The factorization theorem for the differential cross section in electron-positron annihilation is identical to the factorization theorem for measured jet broadening\footnote{Jet broadening, $e$ is defined as $e= \left(\sum_i \vert \vebv{p}^{i} \vert \right)/\omega $ where the transverse momentum is measured with respect to the thrust axis. } derived in Eq.(6.22) of Ref.~\cite{Chiu:2012ir},
\begin{multline}
  \label{eq:SCET2}
  \frac{d\sigma^{\text{(I)}}}{dE_{\perp}}=\sigma_0 H_2 \times \int dE_n dE_{\bar{n}}dE_s \; \delta(E_{\perp} -E_n -E_{\bar{n}}-E_s)
  \int d \vebv{q}^2 d \vebv{p}^2  S(E_s,\vebv{p}^2,\vebv{q}^2)\\\times J(E_n,\vebv{p}^2) \times J(E_{\bar{n}},\vebv{q}^2).
\end{multline}
where $\sigma_0$ is the Born cross section and $H_2$ is the di-jet hard function extracted from the matching of QCD onto SCET and can be found in Refs.~\cite{Bauer:2003di,Manohar:2003vb}. In this factorization theorem the collinear radiation is described through the jet functions, $J(E_{n},\vebv{p}^2)$, where the transverse momentum dependence (measured with respect to the thrust axis) is necessary in order to account for the recoiling of collinear radiation against soft radiation. The contribution of soft radiation is incorporated via the soft function, $S(E_{s},\vebv{p}^2,\vebv{q}^2)$. In this region both collinear and soft radiation contribute the measurement of transverse energy. The relevant modes of SCET that contribute to the factorization theorem in Eq.(\ref{eq:SCET2}) which are the standard SCET$_{\text{II}}$ modes are presented in Table~\ref{tb:modes}.

We now turn our attention to region II where Type-II events dominate the cross section. In order to identify the relevant SCET modes that participate in the factorization of the cross section in this region is important to realize that, in contrast with region I, the collinear modes within the cones outside the veto could have transverse momentum parametrically larger than the soft radiation. This corresponds to approximately back-to-back di-jet events and for this reason we employ SCET$_{\text{I}}$ modes. Furthermore in the limit $r \ll 1$ we need to include the soft and collinear modes that contribute to the measurement and can also resolve the cone boundary. We achieve this using the framework developed in Ref.~\cite{Chien:2015cka} and refactorize the soft function into global soft and collinear-soft functions. Due to the nature of the measurement the two modes have the same virtuality but live in different rapidity regions, as shown in Table \ref{tb:modes}. Thus the factorization of the cross section in region II is,
\begin{equation}
  \label{eq:SCET1}
  \frac{d\sigma^{(\text{II})}}{d \prp{E}} = \sigma_{0} H_{2} \times  J_{q}(\omega) \times J_{\bar{q}}(\omega) \times
 \int dE_n dE_{\bar{n}} \; S_{s}(E_{\perp}-E_{sn}-E_{s\bar{n}}) S_{n}(E_{sn}) S_{\bar{n}}(E_{s\bar{n}}), 
\end{equation}
where $S_s$ is the global soft function and $S_n$ and $S_{\bar{n}}$ are the collinear soft functions associated with corresponding modes.
\tabulinesep=1.7mm
\begin{table}[h!]
  \begin{center}
\begin{tabu}{|c|c|c|}
\hline
 modes              & Region I                             & Region II  \\\hline \hline
 $n$-collinear      & $(\prp{E}^2/\omega,\omega,\prp{E} ) $  & $(\omega r^2,\omega,\omega r )$ \\ \hline
 $n$-collinear-soft & --                                     & $ (\prp{E} r, \prp{E}/r ,\prp{E})$ \\ \hline
 soft               & $(\prp{E},\prp{E},\prp{E}) $           & $(\prp{E},\prp{E},\prp{E})$ \\ \hline
\end{tabu}
 \caption{The scaling of $(p^+,p^-,\prp{p})_n$ for the collinear, soft, and collinear-soft modes for transverse energy measurement where $\hat{n}$ is the thrust axis in electron-positron annihilation, for region I and II.}
  \label{tb:modes}
\end{center}
\end{table}

In fixed order calculations the global soft and collinear-soft functions suffer from rapidity divergences that are regulated with the use of rapidity regulator of  Refs.~\cite{Chiu:2012ir,Chiu:2011qc}. Employing rapidity renormalization group  allows us to resum global logarithms of the cone size parameter $r$ (or equivalently $\ln(1/r) = \vebc{\eta}$). A similar analysis was performed within the context of transverse momentum dependent fragmenting jet functions (TMDFJF)  in Ref.~\cite{Bain:2016rrv}. The jet functions $J_{i}(\omega)$ describe the cone-type ``unmeasured'' jets and can be found  for both gluon, $i=g$, and quark (antiquark), $i=q(\bar{q})$ jets in Ref.~\cite{Ellis:2010rwa}. For large values of $r$ ($r \sim 1$  or equivalently $\vebc{\eta} \to 0$) the factorization theorem in Eq.(\ref{eq:SCET1}) still holds but the refactorization of the soft function is redundant since the collinear-soft and global-soft modes merge to the standard ultra-soft modes of SCET$_{\text{I}}$.  


\subsection{Fixed order results }

In this section we provide the fixed order  results for region II up to NLO accuracy. The elements for resummed expressions up to NLL' are given in the following section along with numerical implementations and comparison with \textsc{Pythia}. For region I all necessary results for the  NLO cross-section  are derived in Ref.~\cite{Chiu:2012ir} and summarized in Appendix~\ref{app:A}.

In region II collinear radiation is contained within the unmeasured cones and therefore does not contribute to the measurement of the transverse energy in the vetoing area. The unmeasured jet function that appears in the factorized expression in Eq.(\ref{eq:SCET1}) are evaluated in Ref.~\cite{Ellis:2010rwa} and are given by:
\begin{equation}
  \label{eq:jet}
J_{q}^{\text{NLO}}(\omega,r)\;\;= 1+
\frac{\alpha_s C_F}{2\pi} \lbc  \frac{7}{2} -\frac{5}{12} \pi^2 + 3\ln(2) +\frac{3}{2}\ln \left(\frac{\mu^2}{r^2 \omega^2} \right)+ \frac{1}{2}\ln^2 \left(\frac{\mu^2}{r^2 \omega^2} \right)\rbc.
\end{equation}
The soft function $S_s(E_s)$ can be calculated using the rapidity and dimensional regulators from the real gluon emission diagrams. Virtual gluon diagrams will give scaleless integrals and therefore are ignored during this calculation. We evaluate the NLO contributions by extending the phase-space integration in Eq.(5.11) of Ref.~\cite{Ellis:2010rwa} with the replacement $\Theta_{\text{alg}} \to \delta(\prp{E}-\prp{k})$ and multiplying by the rapidity regulator factor, $w^2 (\nu / \vert 2  k_{3} \vert) ^{\eta}$,
\begin{align}
  \label{eq:softB}
  S_{s, ij}^{\text{b},(1)}(E_s)&=-g^2 w^2 \left(\frac{e^{\gamma_E}\mu^2}{4 \pi}\right)^{\epsilon}  (\bmat{T}_{i} \cdot \bmat{T}_{j}) \int\frac{dk^+ dk^- d^{d-2}k_{\perp}}{(2\pi)^{d-1}}\; \frac{\nu^{ \eta}  \delta(k^2)}{k^+ k^{-} \vert k^- - k^{+} \vert^{\eta}} \delta(E_s - k_{\perp}) \nn\\
  &= - \frac{\alpha_s w^2}{2 \pi} (\bmat{T}_{i}\cdot \bmat{T}_{j}) \frac{4 e^{ \epsilon \gamma_E}}{\Gamma(1- \epsilon)} \left( \frac{\nu}{\mu}\right)^{\eta} \frac{\Gamma(-\eta)\Gamma(\eta/2)}{\Gamma(-\eta/2)} \frac{1}{\mu}\left( \frac{\mu}{E_s} \right)^{1+2 \epsilon + \eta}.
\end{align}
where the superscript b denotes ``bare'' quantities. Expanding in $\eta$, then in $\epsilon$, and adding the LO contributions we have,
\begin{multline}
S_{s,q \bar{q}}^{\text{b},\text{NLO}} (E_s)=  \delta(E_s)+ \sum_{\substack{i,j=\{q,\bar{q}\}\\ i\neq j}} S_{s, ij}^{\text{b},(1)}(E_s)  =  \delta(E_s)+\frac{\alpha_s w^2 C_F}{ \pi} 
\Big{\lbrace} 
\frac{2}{\eta}\Big{\lbrack}-\frac{1}{\epsilon}\delta(E_s)+2{\cal{L}}_0(E_s,\mu) \Big{\rbrack} \nn \\
+\delta(E_s)\Big{\lbrack}\frac{1}{\epsilon^2}+\frac{1}{\epsilon}\ln \left(\frac{\mu^2}{\nu^2}\right) \Big{\rbrack} - 2{\cal{L}}_0(E_s,\mu)\ln \left(\frac{\mu^2}{\nu^2}\right)-4{\cal{L}}_1(E_s,\mu)-\frac{\pi^2}{12}\delta(E_s)
\Big{\rbrace},
\end{multline}
where we used $\bmat{T}_{q} \cdot  \bmat{T}_{\bar{q}}+\bmat{T}_{\bar{q}} \cdot  \bmat{T}_{q} = - 2 C_F $ and defined 
\footnote{A precise definition of the plus-distributions in Eq.~(\ref{eq:plus}) can be found in Ref.~\cite{Ligeti:2008ac}.}
\begin{equation}
\label{eq:plus}
  \mathcal{L}_n(x,\mu)\equiv \frac{1}{\mu} \mathcal{L}_n \left( \frac{x}{\mu} \right) = \frac{1}{\mu} \lb \frac{\mu}{x} \ln^{n}(x/\mu) \rb_+.
\end{equation}
In the $\overline{\text{MS}}$ scheme, the renormalized result is
\begin{equation}
S_{s,q\bar{q}}^{\text{NLO}} (E_s,\mu,\nu)=  \delta(E_s)- \frac{\alpha_s  C_F}{ \pi} \lbc  2{\cal{L}}_0(E_s,\mu)\ln \left(\frac{\mu^2}{\nu^2}\right)+4{\cal{L}}_1(E_s,\mu)+\frac{\pi^2}{12}\delta(E_s) \rbc,
\end{equation}
where for a generic function, $F$, the bare and renormalized functions are related through the following equation, 
\begin{equation}
  F^{\text{b}}(E)=\int dE \; Z_F(E-E') F(E') \equiv Z_F\otimes F(E),
\end{equation}
and
\begin{equation}
  \label{eq:Zss}
  Z_{ss,q\bar{q}}(E_s)=\delta(E_s)+\frac{\alpha_s w^2 C_F}{ \pi} 
\Big{\lbrace} 
\frac{2}{\eta}\Big{\lbrack}-\frac{1}{\epsilon}\delta(E_s)+2{\cal{L}}_0(E_s,\mu) \Big{\rbrack} + \delta(E_s)\Big{\lbrack}\frac{1}{\epsilon^2}+\frac{1}{\epsilon}\ln \left(\frac{\mu^2}{\nu^2}\right) \Big{\rbrack} \rbc.
\end{equation}
In the renormalized result we take $w \to 1$ but we keep the explicit dependence of the bookkeeping parameter in the renormalization function, $Z_{ss}$, since this will allow us to evaluate the anomalous dimension for the rapidity renormalization group (RRG) in the next section.  

Similarly for the $n$-collinear-soft function, $S_{n}(E_n)$ we have,
\begin{align}
  \label{eq:CSB}
  S_{n, i}^{\text{b},(1)}(E_n) =& 2g^2 w^2 \left(\frac{e^{\gamma_E}\mu^2}{4 \pi}\right)^{\epsilon} \; (\bmat{T}_i)^2 \int\frac{dk^+ dk^- d^{d-2}k_{\perp}}{(2\pi)^{d-1}}\; \frac{\nu^{ \eta}}{k^+ (k^-) ^{1+\eta}} \delta(k^2) \delta(E_n - k_{\perp}) \Theta_{r} \nn \\
  =& - \frac{\alpha_s w^2}{2\pi} (\bmat{T}_i)^2 \frac{4e^{\epsilon\gamma_E }}{\Gamma(1-\epsilon)}\frac{1}{\eta} \left( \frac{\nu r}{\mu} \right)^{\eta} \frac{1}{\mu}\left(\frac{\mu}{E_n}\right)^{1+2\epsilon+\eta},
\end{align}
where $\Theta_r = \Theta(k^{+}/k^{-} - r^2 )$ constrains the collinear-soft gluons to be within the measured region. Following similar process as for the global soft function, we have for the renormalized collinear soft function,
\begin{equation}
  \label{eq:CS}
  S_{n,i}^{\text{NLO}} (E_n,\mu,\nu)=  \delta(E_n) + \frac{\alpha_s  C_i}{2\pi} \lbc  2{\cal{L}}_0(E_n,\mu)\ln \left(\frac{\mu^2}{r^2 \nu^2}\right)+4{\cal{L}}_1(E_n,\mu)+\frac{\pi^2}{12}\delta(E_n) \rbc,
\end{equation}
where $C_{\bar{q}}=C_q \equiv (\bmat{T}_q)^2 = C_F$ (also $C_{g}=C_A$) and the renormalization function is
\begin{equation}
  \label{eq:Zsn}
  Z_{sn,i}(E_n)=\delta(E_n)-\frac{\alpha_s w^2 C_i}{2 \pi} 
\Big{\lbrace} 
\frac{2}{\eta}\Big{\lbrack}-\frac{1}{\epsilon}\delta(E_n)+2{\cal{L}}_0(E_{n},\mu) \Big{\rbrack} + \delta(E_n)\Big{\lbrack}\frac{1}{\epsilon^2}+\frac{1}{\epsilon}\ln \left(\frac{\mu^2}{r^2\nu^2}\right) \Big{\rbrack} \rbc.
\end{equation}


\subsection{Renormalization group evolution and numerics}
\label{sec:2.2}
In this section we review the results from literature regarding the NLL cross section for region I and later study the renormalization properties of the global and collinear-soft functions for the region II. We also perform numerical applications and compare against \textsc{Pythia}.


\subsubsection{Region I NLL cross section}
\label{sec:NLL(I)}

In Ref.~\cite{Chiu:2012ir}   a factorization theorem  is studied for electron-positron annihilation processes where jet broadening, $e$, is measured. This measurement can be related with the measurement of transverse energy in region I through a simple rescaling relation:
\begin{equation}
  e=\frac{1}{\omega} \sum_i \vert \vebv{p}^i \vert= \frac{ \prp{E}}{\omega}.
\end{equation}
The explicit NLL cross section for the simultaneous measurement of the left broadening, $e_L$, and right broadening, $e_R$, is given in Eq.(6.59) of Ref.~\cite{Chiu:2012ir}.
\begin{equation}
\frac{d\sigma^{\text{NLL}}}{de_R de_L} = \sigma_0 U_H (\mu_h,\mu_s) \left(\frac{ e^{\gamma_E} \mu}{\omega} \right)^{-2 \omega_s} \frac{1}{\Gamma^2(\omega_s)\;(e_L e_R)^{1-\omega_s}}  \lb 1- \frac{\omega_s} {2^{-\omega_s}} B_{1/2}(1+\omega_s,0) \rb^2\;,
\end{equation}
where
\begin{align}
\omega_s &= 2 \frac{\alpha_s(\mu)C_F}{\pi} \ln \left( \frac{\nu}{\nu_0} \right), &\text{and} \;\;\;\;\;\;\; B_z(a,b)&=\int_0^z dx\;(1-x)^{b-1}x^{a-1},
  \end{align}
where $\nu$ and $\nu_0$ are the jet and soft rapidity-scales, respectively, and  $B_z(a,b)$ is the incomplete beta-function. Using the fact that jet broadening is the sum of left and right broadening, i.e., $e=e_L+e_R$, we have, 
\begin{equation}
  \frac{d \sigma}{d \prp{E}} (\prp{E})= \int de_L de_R\;\delta(\prp{E}- \omega (e_L +e_R)) \frac{d \sigma }{d e_L de_R}(e_L ,e_R). 
\end{equation}
Performing the integrations we find,
\begin{equation}
  \label{eq:et1}
\frac{d\sigma^{\text{(I), NLL}}}{d\prp{E}} = \sigma_0 U_H (\mu_h,\mu_s) \frac{ e^{-2 \omega_s\gamma_E}}{\Gamma(2\omega_s)}  \frac{1}{\mu} \left(\frac{\mu}{\prp{E}} \right)^{1-2\omega_s}  \lb 1- \frac{\omega_s} {2^{-\omega_s}} B_{1/2}(1+\omega_s,0) \rb^2.
\end{equation}
To determine the jet and soft rapidity scales we look at the fixed order results in Appendix~\ref{app:A1}. We see that at NLO  the rapidity logarithms are minimized with the choices $\nu=\nu_J = \omega$ and $\nu_0=\nu_S = \prp{E}$. Thus for this choice of rapidity, 
\begin{equation}
  \omega_s = 2 \frac{\alpha_s(\mu)C_F}{\pi} \ln \left( \frac{\omega}{\prp{E}} \right).
\end{equation}
We will compare our result for the NLL integrated cross section to  \textsc{Pythia} simulations in the next section.


\subsubsection{Region II NLL' cross section}

The renormalized global and collinear soft functions satisfy the following RG and RRG equations:
\begin{align}
  \label{eq:RG}
  \frac{d}{d \ln(\mu)} F(E,\mu,\nu) &= \gamma^{F}_{\mu}  F(E,\mu,\nu), &
  \frac{d}{d \ln(\nu)} F(E,\mu,\nu) &= \gamma^{F}_{\nu} \otimes F(E,\mu,\nu),
\end{align}
where $\gamma^{F}_{\mu}$ and $\gamma^{F}_{\nu}$ are the RG and RRG anomalous dimensions respectively and $F$ can be either the global soft function, $S_{s,ij}$ or the collinear-soft function $S_{n,i}$. The anomalous dimensions are related to the renormalization function through the following relations\footnote{In the following equations $Z_{F}^{-1}$ is defined such that $Z_{F}^{-1} \otimes Z_{F} (E) = \delta (E)$.}:
\begin{align}
  \gamma^{F}_{\mu}(\mu,\nu) \delta(E) &= - Z_{F}^{-1} \otimes \left( \frac{dZ_{F}}{d\ln(\mu)} \right),  &
  \gamma^{F}_{\nu}(E,\mu)  &= - Z_{F}^{-1} \otimes \left( \frac{dZ_{F}}{d\ln(\nu)} \right),
\end{align}
and thus from Eqs.(\ref{eq:Zss}) and (\ref{eq:Zsn}) we have
\begin{align}
  \label{eq:AD1}
  \gamma_{\mu}^{ss}(\mu,\nu) & = +2\frac{\alpha_s C_F}{\pi} \ln \left( \frac{\mu^2}{\nu^2 } \right), &  \gamma_{\nu}^{ss}(E,\mu) & = +4\frac{\alpha_s C_F}{\pi} \mathcal{L}_0(E,\mu), \nn \\
   \gamma_{\mu}^{sn}(\mu,\nu) & = -\frac{\alpha_s C_F}{\pi} \ln \left( \frac{\mu^2}{\nu^2 r^2} \right), &  \gamma_{\nu}^{sn}(E,\mu) & = -2\frac{\alpha_s C_F}{\pi} \mathcal{L}_0(E,\mu),
\end{align}
in agreement with Eq.(A.9) of Ref.~\cite{Kolodrubetz:2016dzb}. We note that the soft anomalous dimensions satisfy the following consistency relations,
\begin{align}
  \label{eq:cons1}
  \gamma_{\mu}^{ss}(\mu,\nu) +2\;  \gamma_{\mu}^{sn}(\mu,\nu) & =  \gamma_{S}^{\text{unmeas}}(\mu)    & \gamma_{\nu}^{ss}(E,\mu) +2  \; \gamma_{\nu}^{sn}(E,\mu) &= 0,
  \end{align}
where $\gamma_{S}^{\text{unmeas}}(\mu)$ is the soft anomalous dimension from Eq.(6.30) in Ref.~\cite{Ellis:2010rwa}. These consistency  relations are required so that the cross section is independent of the  factorization and rapidity scale. The RRG  anomalous dimension assumes the following generic form to all orders in perturbation theory:
\begin{equation}
 \gamma_{\nu}^{F}(E,\mu) = 2 \Gamma_{\nu}^{F}[\alpha] \mathcal{L}_0(E,\mu) + \Delta\gamma_{\nu}^F[\alpha] \delta(E),
\end{equation}
where $\Gamma_{\nu}^{F}$ is proportional to the cusp anomalous dimension $\Gamma_{\text{cusp}}$ (see Eq.(\ref{eq:G})) and $\Delta\gamma_{\nu}^{F}$ is the non-cusp part of the anomalous dimension.  The values of $\Gamma_{\nu}^F[\alpha]$ and $\Delta\gamma_{\nu}^{F}[\alpha]$ for the collinear and global soft functions for the process $e^+ e^- \to \text{di-jets}$ are given in Table~\ref{tb:Gammas}. 

\begin{table}[h]
  \begin{center}
\begin{tabu}{|c|c|c|c|}
\hline
Function & $\Gamma_{\nu}^F$ & $\Delta\gamma_{\nu}^F$ & $\nu_F$\\
\hline \hline
$ S_{s,q\bar{q}} $ & $2 \alpha_s C_{F} /\pi $ & $\mathcal{O}(\alpha_S^2)$ & $ \prp{E}    $ \\ \hline
$ S_{n,q} $ & $- \alpha_s C_{F} /\pi $ & $\mathcal{O}(\alpha_S^2)$ & $ \prp{E} /r $ \\ \hline
\end{tabu}
 \caption{The rapidity renormalization group anomalous dimensions and rapidity canonical scales, $\nu_F$, for global-soft and collinear-soft  functions for the electron-positron annihilation to di-jets.}
  \label{tb:Gammas}
\end{center}
\end{table}
The solution to the RRGE in  Eq.(\ref{eq:RG}) is:
\begin{equation}
  F(E,\mu,\nu) =  [F(\mu,\nu_F) \otimes \mathcal{V}_F(\mu,\nu,\nu_F)](E),
\end{equation}
where
\begin{equation}
\label{eq:RRGEsoln}
  \mathcal{V}_F(E,\mu,\nu,\nu_F) =  \frac{e^{\kappa_F(\mu,\nu,\nu_F)}(e^{\gamma_E} \mu)^{-\eta_F(\mu,\nu,\nu_F)}}{\Gamma(\eta_F(\mu,\nu,\nu_F))}  \lb \frac{1}{E^{1-\eta_F(\mu,\nu,\nu_F)}} \rb_+,
\end{equation}
where we define the the plus-distribution in Eq.~(\ref{eq:RRGEsoln}) through its inverse Laplace transform,\footnote{For an alternative definition see Appendix C of Ref.~\cite{Fleming:2007xt}.}
\bea
  \lb \frac{1}{E^{1-\alpha}} \rb_+ =  {\cal L}^{-1}\lb s^\alpha \, \Gamma[-\alpha]\rb \nn \; ,
\eea
and 
\begin{align}
  \eta_F(\mu,\nu,\nu_F) &= 2 \Gamma_{\nu}^F[\alpha] \ln \left( \frac{\nu}{\nu_F}  \right),  &  \kappa_F(\mu,\nu,\nu_F) &=  \Delta\gamma_{\nu}^F[\alpha] \ln \left( \frac{\nu}{\nu_F} \right),
\end{align}
where $\nu_F$ is the characteristic scale from which we start the evolution and is chosen such that at this scale rapidity logarithms are minimized. For the global and collinear soft functions these are given in Table~\ref{tb:Gammas}. The solution of the RGE in Eq.(\ref{eq:RG}) has been described previously in the literature (see for example Ref.~\cite{Ellis:2010rwa}) and is summarized in Appendix~\ref{app:A}. 

\begin{figure}[t!]
  \centerline{\includegraphics[width = \textwidth]{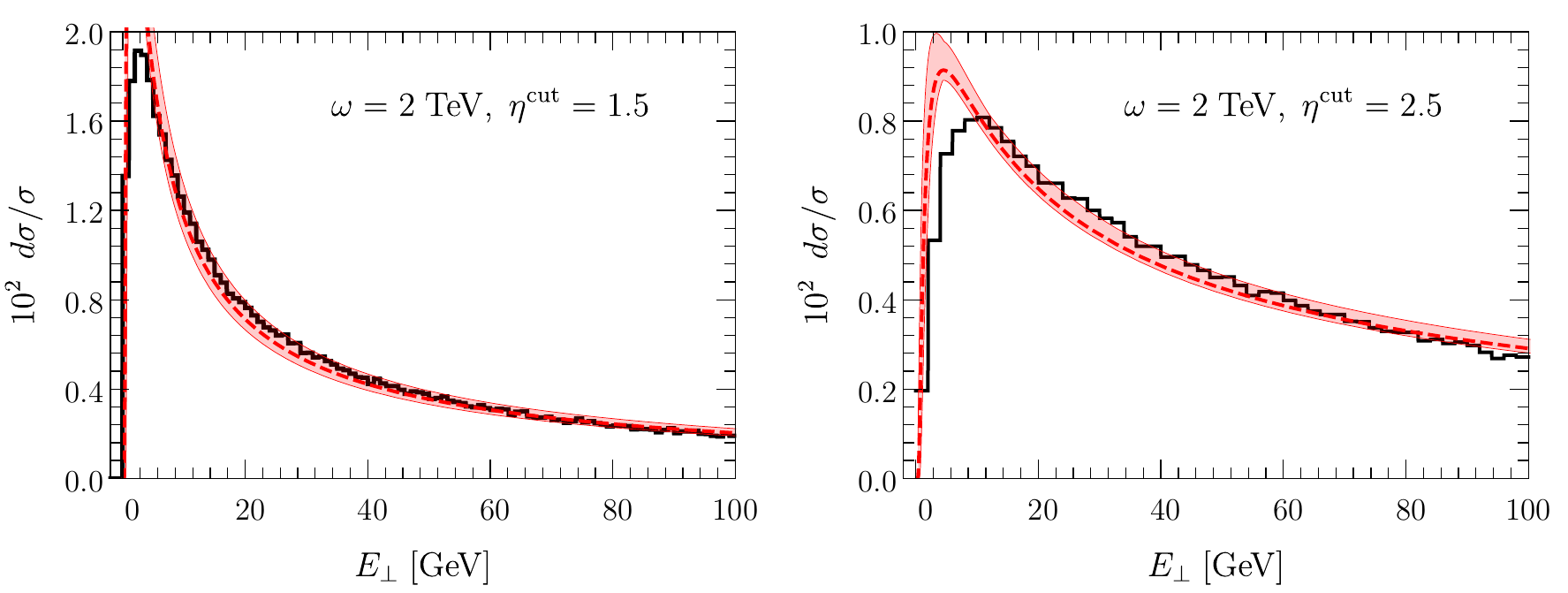}}
\caption{The differential cross section as a function of the transverse energy from \textsc{Pythia} simulation (black solid) against NLL' analytic calculations (red band) for region II for $\omega= 2$ TeV and for $\vebc{\eta}=1.5$ (left) where $\omega r \sim 450$ GeV and $\vebc{\eta}=2.5$ (right) where $\omega r \sim 164$ GeV. The simulation data are normalized to unity and the analytic results are normalized to data in the region $\prp{E} < \omega r/2$.}
\label{fig:dig}
\end{figure}

In order to calculate the cross section up to  next-leading-logarithmic (NLL) accuracy we evolve the hard and the jet function from their characteristic scales ($\mu_H = \omega $ and $\mu_J = \omega r$) to the soft scale, $\mu_{ss} =\mu_{sn} = \prp{E}$, as discussed in Appendix~\ref{app:A.3}. We also perform the evolution in the rapidity space by evolving the global soft function from $\nu_{ss} = \prp{E} $ to the collinear soft function canonical scale, $\nu_{sn} = \prp{E}/r$. The evolution in the rapidity space allow us to resum global logarithms of $r$ up to NLL accuracy. Thus our final result for the NLL' cross section in region II is:
\begin{multline}
  \label{eq:cs2}
  \frac{1}{\sigma_0} \frac{d \sigma^{\text{(II), NLL'}}}{d\prp{E}} = \mathcal{U}_{H}(\mu_{ss},\mu_h) H^{\text{NLO}}_2(\omega,\mu_h) \times (\mathcal{U}_{J}(\mu_{ss},\mu_j) J^{\text{NLO}}_{q}(\omega,r,\mu_j))^2  \times 
  \frac{(e^{\gamma_E} \mu_{ss})^{-\eta_{ss}}}{\Gamma(\eta_{ss})} \\
    \times  \frac{1}{\prp{E}^{1-\eta_{ss}}} \left(1 - 4 \frac{\alpha_s(\mu_{ss})C_F}{\pi} \ln \left( \frac{ \nu_{sn}r}{\nu_{ss}} \right)   f_0(\prp{E},\mu_{ss},\nu_{sn},\nu_{ss}) \right),
\end{multline}
where
\begin{equation}
  \eta_{ss} \equiv \eta_{ss}(\mu_{ss},\nu_{sn},\nu_{ss}),
\end{equation}
and
\begin{equation}
  f_0(E,\mu,\nu_{sn},\nu_{ss}) = \ln \left( \frac{E}{\mu} \right) - H(-1+ \eta_{ss}(\mu,\nu_{sn},\nu_{ss}) ), 
  \end{equation}
where $H(x)$ is the harmonic number function. In the final result the scales are fixed after we performed the convolutions. We emphasize again that in this work we resum only the large global logarithms and resummation of NGLs is beyond the scope of this work. Unfortunately, the cross section in region II suffers from non-global logarithmic enchantments of the form, $\ln(\prp{E}/\omega r)$, and a complete description of the cross section  at NLL or NLL' in region II requires nontrivial  resummation of NGLs which we leave for future work.  Recently resummation of NGLs for a similar observable  was achieved in Ref.~\cite{Becher:2017nof} using the technology of multi-Wilson-line ``coft-functions''.

In Fig.~\ref{fig:dig} we compare our result for the NLL' cross section (red band) in region II against \textsc{Pythia} simulations (back solid line) at $\omega = 2$ TeV for $\vebc{\eta} = 1.5 $ (left) and $\vebc{\eta} = 2.5$ (right). For the simulations hadronization is turned off. The analytic results are normalized to the simulations in the region $0 < \prp{E} < \omega r/2 $. Theoretical uncertainties were estimated by fluctuating  all canonical scales by a factor of 2 and 1/2. We find that within the theoretical uncertainties the NLL'  results agree very well with the simulations for most values of $E_\perp$. There is  disagreement  near the peak of the distribution which could be due to the strong coupling constant in the soft function being evaluated at relatively small scales where higher orders in perturbation theory and higher logarithmic  accuracy might be required for reliable results.

\begin{figure}[t!]
  \centerline{\includegraphics[width = \textwidth]{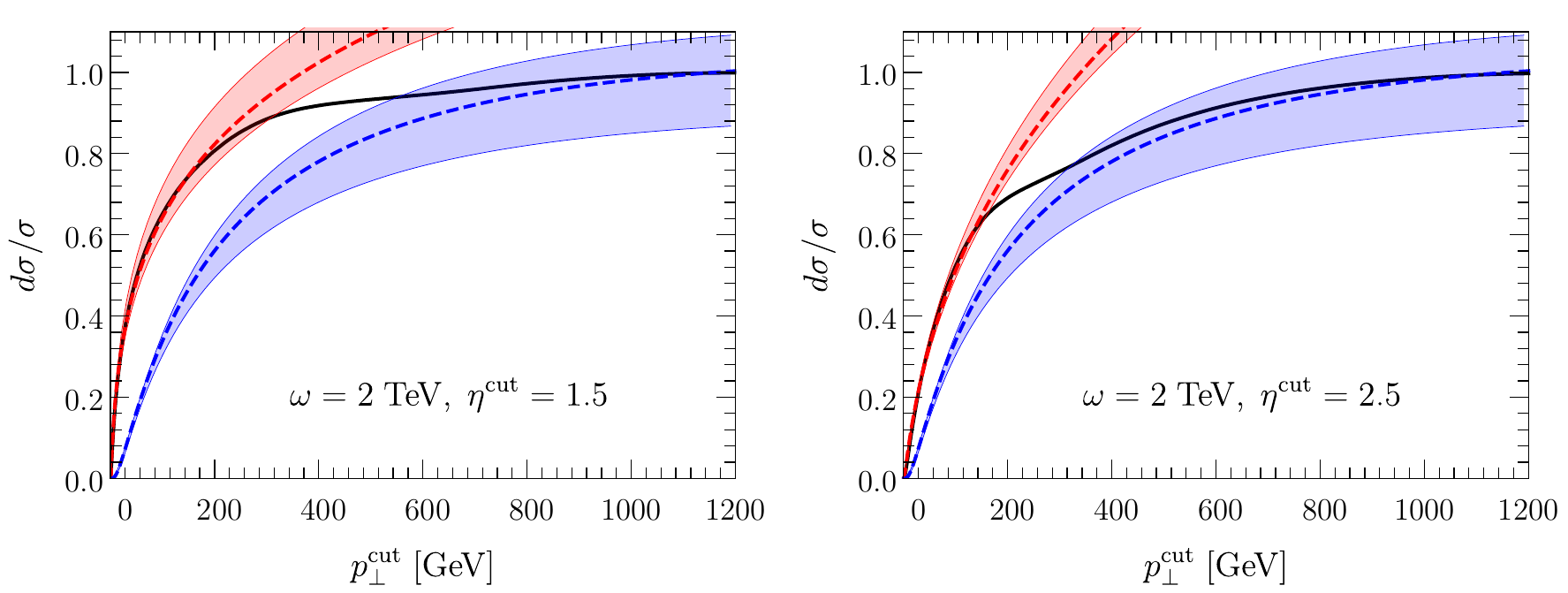}}
\caption{The integrated cross section as a function of the vetoing parameter $\veb{p}$ from \textsc{Pythia} simulation (black solid) compared against the small transverse energy approximation (region II, red band) and the large transverse energy approximation (region I, blue band) for $\omega= 2$ TeV and for $\vebc{\eta}=1.5$ (left) where $\omega r \sim 450$ GeV and $\vebc{\eta}=2.5$ (right) where $\omega r \sim 164$ GeV.}
\label{fig:cum}
\end{figure}

In Fig.~\ref{fig:cum} we compare our result for the integrated cross section (red band) defined in Eq.(\ref{eq:cum}) against \textsc{Pythia} for same kinematic variables. In this figure we also extend the graphs into region I and we include the predictions for that region from the factorization theorem in Eq.(\ref{eq:et1}) for the  NLL cross section (blue band). We see that the at small and large values of $\veb{p}$  the cross section is well described by $d\sigma^{{\text{(II)}}}$ and $d\sigma^{{\text{(I)}}}$ respectively. The intermediate regime where $\veb{p} \sim \omega r$ could be described with the use of a phenomenologically motivated combination of the cross sections that interpolates between the two regions, but developing such a formula  is beyond the scope of this work.  

\subsection{Jet function corrections for $\prp{E} \lesssim \omega r$}

In this section we consider corrections when $\prp{E} \lesssim \omega r$ to the jet function, $J_{q}(\omega,r)$, used in the factorization theorem for the region II cross section. Contributions in this regime come from out-of-jet radiation of collinear modes. In electron-positron annihilation these corrections were discussed in Ref.~\cite{Ellis:2010rwa} for the integrated jet function in the context of energy veto. Here we extend the calculation for measurements of the transverse energy with respect to thrust axis. For this purpose we modify the factorization theorem in Eq.(\ref{eq:SCET1}) such that includes contributions to the measurement from collinear radiation. This modification allows us to extend region II to $\prp{E} \lesssim \omega r \ll \omega$ which we refer to as region II.e. The cross section is given by,
\begin{equation}
  \label{eq:SCET1corr}
  \frac{d\sigma^{(\text{II.e})}}{d \prp{E}} = \sigma_{0} H_{2}  \times S_{s} \otimes [ S_{n} \otimes  \mathcal{J}_{q}(\omega,r)]^2 ,
\end{equation}
where at NLO
\begin{equation}
  \label{eq:jetcorr}
\mathcal{J}_{q}^{\text{NLO}}(E_{\perp},\omega,r)\;\;=\; J_{q}^{\text{NLO}}(\omega,r) \delta(E_{\perp}) + \Delta J_{q}^{(1)} (E_{\perp},\omega,r),
\end{equation}
where
\begin{multline} 
\Delta J_{q}^{(1)} (E_{\perp},\omega,r)\;= -\frac{\alpha_s C_F}{2\pi} \lbc  \Theta \Big{(}E_{\perp} \Big{)} \Theta \Big{(}\frac{\omega r}{2} -E_{\perp} \Big{)} \lb 6 \left(\frac{1}{r \omega} \right) + \frac{4}{E_{\perp}} \ln \left(1-\frac{E_{\perp}}{r \omega} \right) \rb \\
-\Theta \Big{(}E_{\perp}-\frac{\omega r}{2}  \Big{)}  \Theta \Big{(} \omega r - E_{\perp} \Big{)} \lb 6 \left(\frac{1}{r \omega} \right) - \frac{6}{E_{\perp}}+ \frac{4}{E_{\perp}} \ln \left(1-\frac{E_{\perp}}{r \omega} \right) - \frac{8}{E_{\perp}} \ln \left(\frac{E_{\perp}}{r \omega} \right) \rb \rbc.
\end{multline}
The first component of the jet function in Eq.(\ref{eq:jetcorr}) is given by the standard unmeasured jet function in Eq.(\ref{eq:jet}) multiplied by $\delta(\prp{E})$ since the unmeasured jet function is calculated for the case of both partons emitted within the unmeasured cone. The calculation of the corrections from out-of-jet radiation, $\Delta J_{q}^{(1)}$, are summarized in Appendix~\ref{app:A}.  The cross section in this region is then given by
\begin{equation}
  \label{eq:cs2e}
  \frac{d\sigma^{\text{(II.e), NLL'}}}{d \prp{E}} = \frac{d\sigma^{\text{(II), NLL'}}}{d \prp{E}} + 2 \frac{d\sigma^{\text{(II), NLL}}}{d \prp{E}} \otimes \Delta J_{q}^{(1)}.
  \end{equation}

\begin{figure}[t!]
  \centerline{\includegraphics[width = \textwidth]{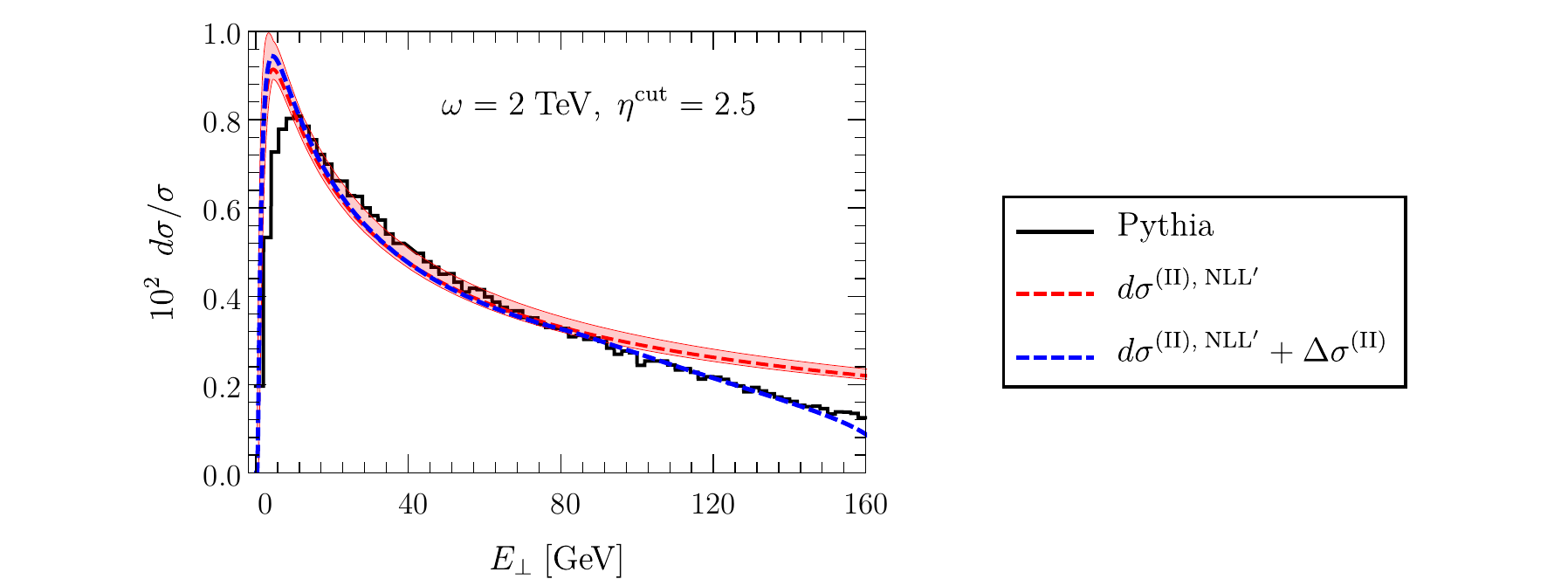}}
\caption{The differential cross section as a function of the transverse energy from \textsc{Pythia} simulation (black solid) against NLL' analytic calculations without jet function corrections (red band) and with jet function corrections (green dashed)  for region II.e for $\omega= 2$ TeV and for $\vebc{\eta}=2.5$.}
\label{fig:outcorr}
\end{figure}

In Fig.~\ref{fig:outcorr} we compare the analytic expression from Eq.(\ref{eq:cs2}) and (\ref{eq:cs2e}) for region II and region II.e respectively against \textsc{Pythia} simulations. We find that for  $\prp{E} < \omega r/2 $, the two expressions give almost identical distributions. This suggests that in this region, the contribution from the out-of-jet radiation to the cross section, denoted $\Delta \sigma^{\text{(II)}}$, is power suppressed, as expected. On the other hand, in the region  $\omega r/2 < \prp{E} < \omega r$ where $\Delta \sigma ^{\text{(II)}}$ is expected to give a significant contribution to the cross section, the distribution for  $d\sigma^{\text{(II.e), NLL'}} = d\sigma^{\text{(II), NLL'}} +\Delta \sigma^{\text{(II)}} $ shows improved agreement with the simulation data compared to $d\sigma^{\text{(II), NLL'}}$. It should be noted that in this region NGLs of the form $\ln(\prp{E}/\omega r)$ are not large and therefore can be included order by order in perturbation theory.

\section{Extension to hadronic collisions }
\label{sec:pp}
In this section we extend the analysis of the previous chapter  to hadronic collisions. We consider the effects of rapidity cutoff in transverse energy and jet-veto measurements where the transverse momentum and rapidity are measured with respect to the beam axis (hence the change of notation $\prp{E} \to \tra{E}$). A significant difference  from the analysis of  electron-positron annihilation is  that for $O(\alpha_s)$ calculations in hadronic collisions, only one parton can contribute to the measurement. Therefore, the hard-interacting (incoming) parton is not constrained by the rapidity cutoff. In direct analogy to the previous section we identify Region II.e as $\tra{E} \lesssim p^{-} r \ll p^{-}$ where $p^{-}=x_{B}E_{\text{cm}}$ is the large component of the lightcone momenta of the incoming parton and $E_{\text{cm}}$ is the hadronic center of mass energy. Like before we define  $r=e^{-\vebc{\eta}}$ where $\vebc{\eta}$ is the rapidity cutoff.

We focus primarily on  the  factorization theorem for the di-jet cross section but these results can be  straightforwardly extended to the case of zero and one jet cross sections.  For  measured transverse energy in SCET the factorized cross section is,
\begin{equation}
    \frac{d\sigma^{\text{(II.e)}, ab \to 12}}{dy_1 dy_2 dp_{T} d\tra{E}}  = \mathcal{N}  \mathcal{B}_{a/P}(x_1,\mu) \otimes \mathcal{B}_{b/P}(x_2,\mu)
     \otimes \Tr [\bmat{H}^{ab}_{12}(\mu)\bmat{S}^{ab}_{12}(\tra{E},\mu)]\; J_1(\mu) \; J_2(\mu),
\end{equation}
where
\begin{equation}
  \mathcal{N}\equiv\frac{p_T}{32 \pi N_{c}  x_1 x_2E_{\text{cm}}^4}.
  \end{equation}
The corresponding factorization theorem for jet veto measurements is discussed in Section~\ref{sec:app}. In this section we will evaluate the NLO di-jet refactorized soft function for transverse energy, $\bmat{S}^{ab}_{12}(\tra{E},\mu)$, and we discuss how from this result we can trivially construct  the corresponding soft function for  jet-veto measurements, $\bmat{S}^{ab}_{12}(\veb{p},\mu)$. The refactorized result involves the global soft function for which no rapidity cutoff is implemented and soft radiation is allowed within the jet cones. For zero and one jet production the global soft function for is given at NLO in Ref.~\cite{Kolodrubetz:2016dzb}.  The missing elements for the refactorized soft function are the beam and jet collinear-soft function which we also present in this section and discuss their evolution properties. For our calculations we assume that the jet and beam directions are well separated. 

Furthermore we give all perturbatively calculable elements for constructing the unmeasured quark beam function and give corrections to those from the out-of-beam radiation.  The details of this calculation are given in Appendix~\ref{app:B} and in this section we summarize our results. 


\subsection{The di-jet soft function}

At NLO the soft function can be written in the following form (see Refs.\cite{Hornig:2016ahz,Ellis:2010rwa})
\begin{align}
  \bmat{S}(\tra{E},\{\tau\})&= \bmat{S}_{\text{unmeas}}(\tra{E})\prod_{i}^{m} S_{\text{meas}}(\tau^{(i)}) \nn  \\
  &=\bmat{S}^{\text{NLO}}_{\text{unmeas}}(\tra{E})\prod_{i}^{m} \delta(\tau^{(i)}) +  \bmat{S}_0 \lb \sum_i^{m} S^{(1)}_{\text{meas}}(\tau^{(i)}) \prod_{k\neq i}^{m} \delta(\tau^{(k)}) \rb\delta(\tra{E}),
  \end{align}
where $m$ can be either 0,1, or 2 and is the number of jets for which the jet substructure observable, $\tau$, is measured,  $\bmat{S}_0$ is the tree-level soft function, which can be found in  Refs.~\cite{Hornig:2016ahz,Kelley:2010fn}, and $\bmat{S}_{\text{unmeas}}$ is the universal di-jet soft function which is independent of the type of jet substructure measurement performed within the jets. Here bold-faced indicates quantities that are matrices in color space. The only non-trivial function of the measurements is $S_{\text{meas}}(\tau)$ which describes the contribution to the jet-shape measurement from soft radiation within the jet. It should be noted that for the completely unmeasured case ($m=0$) the soft function reduces to $\bmat{S}_{\unmeas}$. In this section we will evaluate $\bmat{S}_{\text{unmeas}}$ and discuss its renormalization group properties.

The universal part of the soft function recieves contributions from the global-soft and collinear-soft modes along the direction of beam and jets. Thus $\bmat{S}_{\text{unmeas}}$ is factorized into five terms:
\begin{equation}
  \bmat{S}_{\text{unmeas}}(\tra{E}) = \bmat{S}_s \otimes S_{n,B} \otimes S_{n,\overbar{B}} \otimes S_{n,1} \otimes S_{n,2}.
  \end{equation}
We construct the di-jet global soft function, $\bmat{S}_s$, for transverse energy and jet veto measurements using results from the literature and we calculate $S_{n,i}$. For this reason we organize this calculation in similar way as in Ref.~\cite{Hornig:2016ahz} where the authors calculated the universal (unmeasured) part of the di-jet soft function in the limit $r \to 1$. In contrast with the calculation in Ref.~\cite{Hornig:2016ahz} in our calculation of the global soft function, gluons from real emissions could have pseudo-rapidity greater than $\vebc{\eta}$, thus:
\begin{equation}
  \label{eq:ppsoft}
  \bmat{S}^{\text{b,NLO}}_s = \bmat{S}_0 + \lb \bmat{S}_0 \sum_{i<j} (\bmat{T}_i\cdot\bmat{T}_j) S^{(1)}_{ij} + \text{h.c.} \rb,
\end{equation}
where $S_{ij}^{(1)}$ is the inclusive $i$-$j$ interference term,
\begin{equation}
  S_{ij}^{(1)} \equiv  -g^2 \lp \frac{e^{\gamma_E} \mu^2}{4 \pi}  \rp^{\epsilon} w^2\int \frac{d^d k}{(2\pi)^{d-1}} \frac{n_i \cdot n_j}{(n_i \cdot k)(n_j \cdot k)}\lp\frac{\nu}{2 k_L}\rp^{\eta} \delta(k^2) \delta(\tra{E}- \tra{E}(\bmat{k})) \Theta(k_0). 
\end{equation}
where $\tra{E} (\bmat{k})$ is the transverse energy as a function of the gluon three-momenta, $\bmat{k}$, and $k_L$ is longitudinal component  with respect to the beam axis.

Since in this paper we are considering hadronic collisions in the center of mass frame where $\hat{n}_{B}$ and $\hat{n}_{\overbar{B}}$ are always back to back, the beam-beam interference term, $S^{(1)}_{B\overbar{B}}$ is given by Eq.(\ref{eq:softB}) up to the color factor which is factored out of Eq.(\ref{eq:ppsoft}), so
\begin{multline}
S_{B\overbar{B}}^{(1)} (\tra{E})= \frac{\alpha_s(\mu) w^2}{ 2\pi} 
\Big{\lbrace} 
\frac{2}{\eta}\Big{\lbrack}\frac{1}{\epsilon}\delta(\tra{E})-2{\cal{L}}_0(\tra{E},\mu) \Big{\rbrack}  \\
-\delta(\tra{E})\Big{\lbrack}\frac{1}{\epsilon^2}-\frac{2}{\epsilon}\ln \left(\frac{\nu}{\mu}\right) \Big{\rbrack} +4{\cal{L}}_1(\tra{E},\mu)- 4{\cal{L}}_0(\tra{E},\mu)\ln \left(\frac{\nu}{\mu}\right)+\frac{\pi^2}{12}\delta(\tra{E})
\Big{\rbrace}.
\end{multline}
The beam-jet interference term, $S^{(1)}_{BJ}$, where $J$ is either 1 or 2, is one-half times the sum of Eqs.(B.2) and (B.10) of Ref.~\cite{Kolodrubetz:2016dzb},
\begin{align}
  S^{(1)}_{BJ}(\tra{E})= & \frac{\alpha_s(\mu) w^2}{2 \pi} \lbc  \frac{1}{\eta} \lb \frac{1}{\epsilon} \delta(\tra{E}) - 2 \mathcal{L}_0(\tra{E},\mu)\rb + \frac{1}{\epsilon} \mathcal{L}_0(\tra{E},\mu)\nn \\
 & + \frac{1}{\epsilon} \lb \ln\left( \frac{\nu}{\mu} \right) - \frac{1}{\epsilon} - \eta_J \rb \delta(\tra{E})  
   + 2 \mathcal{L}_0(\tra{E},\mu) \lb \eta_J-\ln\left( \frac{\nu}{\mu} \right) \rb +  \frac{\pi^2}{12}\delta(\tra{E})  
  \rbc. 
\end{align}

The jet-jet interference contribution  to the global soft function is given in Table 1 of Ref.~\cite{Hornig:2016ahz} (up to terms that are suppressed by $\mathcal{O}(R^2)$\footnote{It was shown in Ref.~\cite{Bertolini:2017efs} (where the $\mathcal{O}(R^2)$ contributions were studied), even for moderate values of the jet size ($R\lesssim 1$) the small $R$ limit gives a very good approximation to the soft function.}). It should be noted that the jet-jet interference terms do not have rapidity divergences and therefore are independent of the rapidity regulator parameters $\eta$ and $\nu$,
\begin{equation}
  S^{(1)}_{12}(\tra{E}) = \frac{\alpha_s(\mu) w^2}{2 \pi} \frac{1}{\mu} \Big{(} \frac{\mu}{\tra{E}} \Big{)}^{1+2\epsilon} (2 \cosh(\Delta \eta/2))^{-2\epsilon} \lb  \frac{2}{\epsilon} - \epsilon \lp{\Delta \eta}^2 + \frac{\pi^2}{6}\rp\rb,
\end{equation}
where $\Delta \eta = \eta_1-\eta_2$. Expanding in $\epsilon$ and using the two following relations (see Eq.(4.17) of Ref.~\cite{Hornig:2016ahz})
\begin{equation}
  \ln \lp 2 \cosh(\Delta \eta/2) \rp =  \frac{1}{2} \lb \ln \lp \frac{ n_1 \cdot n_2}{2} \rp + \ln( 2 \text{ch}_1) +\ln( 2 \text{ch}_2) \rb\;,
\end{equation}
where $\text{ch}_i \equiv \cosh(\eta_i)$ and
  \begin{equation}
   \ln^2 \lp 2 \cosh(\Delta \eta/2) \rp  = \frac{\Delta \eta^2}{4} + \ln(1+e^{\Delta \eta})\ln(1+e^{-\Delta \eta})\;,
  \end{equation}
 we  get for the jet-jet interference term,
\begin{multline}
  S^{(1)}_{12}(\tra{E}) = \frac{\alpha_s(\mu) w^2}{2 \pi} \lbc
  \lb  \frac{1}{\epsilon} \delta(\tra{E}) - 2 \mathcal{L}_0(\tra{E},\mu)  \rb \lp \ln \lp \frac{ n_1 \cdot n_2}{2} \rp + \ln( 2 \text{ch}_1) +\ln( 2 \text{ch}_2) \rp \\
  - 2 \ln(1+e^{\Delta \eta})\ln(1+e^{-\Delta \eta}) \delta(\tra{E})
  -\frac{1}{\epsilon} \lb  \frac{1}{\epsilon} \delta(\tra{E}) - 2 \mathcal{L}_0(\tra{E},\mu)  \rb - 4 \mathcal{L}_1 (\tra{E},\mu) +\frac{\pi^2}{12}\delta(\tra{E})
  \rbc.
  \end{multline}
 Adding all contributions and using color conservation (i.e. $\bmat{T}_1+\bmat{T}_2+\bmat{T}_B+\bmat{T}_{\bar{B}}=0$) to simplify the results we find, 
\begin{multline}
  \label{eq:ppsoftg}
  \bmat{S}_s^{\text{b,NLO}}(\tra{E})=\bmat{S}_0 \delta(\tra{E})+\frac{\alpha_s(\mu) w^2}{2 \pi} \lb \bmat{S}_0 \lbc \lb \frac{1}{\epsilon} \delta(\tra{E}) - 2 \mathcal{L}_0(\tra{E},\mu)  \rb \Big{(}  \sum_{i} C_i\ln \left( \frac{\bar{\mu}_i}{m_i} \right) -\bmat{M}' (m_i)\Big{)}  \\
   - 2(\bmat{T}_1 \cdot \bmat{T}_2) \ln(1+e^{\Delta \eta})\ln(1+e^{-\Delta \eta})  \delta(\tra{E})  \\
   + (C_1+C_2) \Big{(} \frac{1}{2\epsilon}\lb \frac{1}{\epsilon}\delta(\tra{E})- 2\mathcal{L}_0(\tra{E},\mu)\rb+2\mathcal{L}_1(\tra{E},\mu) -\frac{\pi^2}{24} \delta(\tra{E})  \Big{)}\\
   -( C_B +C_{\overbar{B}})\Big{(}\frac{1}{\eta} \lb \frac{1}{\epsilon} \delta(\tra{E}) -2\mathcal{L}_0(\tra{E},\mu) \rb 
  +\frac{1}{2 \epsilon} \lb 2\ln\left(\frac{\nu}{\mu}\right) - \frac{1}{\epsilon}\rb  \delta(\tra{E}) \\
   -2 \mathcal{L}_0(\tra{E},\mu) \ln\left(\frac{\nu}{\mu}\right) + 2\mathcal{L}_1(\tra{E},\mu) + \frac{\pi^2}{24}  \delta(\tra{E})
  \Big{)} \rbc+ \text{h.c.} \rb,
\end{multline}
where $\bar{\mu}_{1,2} = p_{T}$, $\bar{\mu}_{B, \overbar{B}} = x_i E_{\text{cm}}$ \footnote{Note: the scales $\bar{\mu}_i$ are not the canonical scales $\mu_i$ for which logarithms of the factorization scale are minimized in the corresponding functions.} and
\begin{equation}
  \label{eq:Mp}
  \bmat{M}'(m_i)= - \sum_{i<j} (\bmat{T}_i \cdot \bmat{T}_j) \ln \left( \frac{s_{ij}}{m_i m_j} \right),
  \end{equation}
where $s_{ij} \equiv 2 p_i \cdot p_j$ and $p^{\mu}_{1}$ and $p^{\mu}_2$ are the four-momenta of the two jets and $p^{\mu}_{B(\overbar{B})} = (x_{B(\overbar{B})} E_{\text{cm}}/2) n^{\mu}_{B(\overbar{B})}$. The renormalized soft function is defined by
\begin{equation}
  \bmat{S}_s^{\text{b}}=\bmat{Z}_{ss}^{\dag}(\mu) \otimes  \bmat{S}_s (\mu) \otimes \bmat{Z}_{ss}(\mu),
\end{equation}
where $\bmat{S}_s$ is the renormalized global soft function and $\bmat{Z}_{ss}$ is the corresponding renormalization matrix.  The renormalized soft function satisfies the following RG and RRG equations,
\begin{align}
  \label{eq:GSRGE}
  \frac{d}{d\ln(\mu)}\bmat{S}_s(\tra{E}) &= \bmat{S}_{s}  \otimes  \bmat{\Gamma}_{\mu}^{ss}+ \text{h.c.}
  ,&
  \frac{d}{d\ln(\nu)}\bmat{S}_s(\tra{E}) &= \gamma_{\nu}^{ss} \otimes \bmat{S}_{s}, 
  \end{align}
In the $\overline{\text{MS}}$ scheme we have
\begin{multline}
  \label{eq:ppsoftgR}
  \bmat{S}^{\text{NLO}}_s(\tra{E})=\bmat{S}_0 \delta(\tra{E})+\frac{\alpha_s(\mu) }{2 \pi} \lb \bmat{S}_0 \lbc  - 2 \mathcal{L}_0(\tra{E},\mu)  \Big{(}  \sum_{i} C_i\ln \left( \frac{\bar{\mu}_i}{m_i} \right) -\bmat{M}' (m_i)\Big{)}   \\
   - 2(\bmat{T}_1 \cdot \bmat{T}_2) \ln(1+e^{\Delta \eta})\ln(1+e^{-\Delta \eta})  \delta(\tra{E})  + (C_1+C_2) \Big{(}2\mathcal{L}_1(\tra{E},\mu) -\frac{\pi^2}{24} \delta(\tra{E})  \Big{)}  \\
   -( C_B +C_{\overbar{B}})\Big{(} -2 \mathcal{L}_0(\tra{E},\mu) \ln\left(\frac{\nu}{\mu}\right) + 2\mathcal{L}_1(\tra{E},\mu) + \frac{\pi^2}{24}  \delta(\tra{E}) \Big{)}  \rbc+ \text{h.c.} \rb ,
\end{multline}
and the corresponding anomalous dimensions are
\begin{equation}
  \bmat{\Gamma}_{\mu}^{ss} = \frac{\alpha_s}{\pi} \lb \sum_{i} C_i\ln \left( \frac{\bar{\mu}_i}{m_i} \right)-\bmat{M}'\rb\delta(\tra{E})+\frac{1}{2}\gamma_{\mu}^{ss} (\tra{E})\;,  
\end{equation}
and
\begin{equation}
  \gamma_{\nu}^{ss}= 2 (C_B+C_{\overbar{B}})\frac{\alpha_s}{\pi}\mathcal{L}_0(\tra{E},\mu),
  \end{equation}
where $\gamma_{\mu}^{ss}$ is the total color-trivial part of the soft anomalous dimension after adding the hermitian conjugate in Eq.(\ref{eq:GSRGE}),
\begin{equation}
  \gamma_{\mu}^{ss}(\tra{E})=2\frac{\alpha_s}{\pi} \lb (C_B+C_{\overbar{B}})\ln\Big{(}\frac{\mu}{\nu} \Big{)} \delta(\tra{E})-(C_1+C_2) \mathcal{L}_0 (\tra{E},\mu)\rb.
  \end{equation}

The collinear-soft function $S_{n,i}$ involves collinear-soft fields along the $\hat{n}_i$-direction only, and those fields are decoupled at the level of the effective Lagrangian from the soft Wilson lines along the other $\hat{n}_{i\neq j}$ directions. The one loop contribution to the collinear-soft function along the jet direction ($\hat{n}_J$) is given by:
\begin{align}
  \label{eq:CSJ}
  S_{n,J}^{\text{b},(1)}(\tra{E}) =& 2g^2  \left(\frac{e^{\gamma_E}\mu^2}{4 \pi}\right)^{\epsilon} \; (\bmat{T}_i)^2 \int\frac{dk^+ dk^- d^{d-2}k_{\perp}}{(2\pi)^{d-1}}\; \frac{1}{k^+ k^-} \delta(k^2) \delta \lp \tra{E} - \frac{s_J k^-}{2} \rp \Theta_{R} \nn \\
  =& \frac{\alpha_s C_J}{ 2\pi} \frac{e^{\epsilon \gamma_E}}{\Gamma(1-\epsilon)} \frac{2}{\epsilon R^{2\epsilon}}\frac{1}{\mu} \lp \frac{\mu}{\tra{E}} \rp^{1+2\epsilon} ,
\end{align}
where $s_J = \sin(\theta_{J})$ wth $\theta_J$ the angle between the jet axis and the beam axis, and $\Theta_{R} = \Theta(k^+ / k^- - (s_J R /2)^2)$. Expanding in $\epsilon$  we get,  
\begin{equation}
  S^{\text{b,NLO}}_{n,J} (\tra{E})= \delta(\tra{E}) -\frac{\alpha_s C_J}{ 2\pi} \lbc \frac{1}{\epsilon^2} \delta(\tra{E})- \frac{2}{\epsilon} \mathcal{L}_0(\tra{E},\mu/R)\rbc + S_{n,J}^{(1)} +\mathcal{O}(\epsilon),
\end{equation}
where
\begin{equation}
  \label{eq:CSnJ}
  S_{n,J}^{(1)}(\tra{E})=  -\frac{\alpha_s C_J }{ 2 \pi} \lbc 4\mathcal{L}_0(\tra{E},\mu)   \ln R  + 4\mathcal{L}_1(\tra{E},\mu)+  \Big{(} 2 \ln^2 R -\frac{\pi^2}{12} \Big{)} \delta(\tra{E})  \rbc.
\end{equation}
The renormalized quantity is defined through,
\begin{equation}
  \label{eq:RGsnJ}
 S_{n,J}^{\text{b}} (\tra{E})= Z_{sn,J}(\mu) \otimes S_{n,J}(\tra{E},\mu),
\end{equation}
and satisfies the following RGE
\begin{equation}
  \frac{d}{d \ln (\mu)} S_{n,J}(\tra{E},\mu) = \gamma_{\mu}^{sn,J}(\mu) \otimes S_{n,J}(\tra{E},\mu),
\end{equation}
where $\gamma^{sn,J}_{\mu}$ is the RG anomalous dimension. It should be noted that since the collinear-soft function along the jet direction does not have rapidity divergences, it  does not evolve in the rapidity space. In the $\overline{\text{MS}}$ scheme the renormalized collinear soft function and the corresponding anomalous dimension are,
\begin{align}
   & S^{\text{NLO}}_{n,J} (\tra{E})= \delta(\tra{E}) +  S_{n,J}^{(1)}(\tra{E}),\;\;\;\text{ and } &\gamma^{sn,J}_{\mu}(\tra{E}) = 2 \frac{\alpha_s C_J}{\pi} \mathcal{L}_0(\tra{E},\mu/R),
  \end{align}
respectively. The collinear-soft function along the beam direction, up to the color factor, is identical to the one we evaluated in Eq.(\ref{eq:CS}).
\begin{equation}
  \label{eq:ppsoftn}
  S_{n,B}^{\text{NLO}} (\tra{E})=  \delta(\tra{E}) + \frac{\alpha_s  C_B}{2\pi} \lbc  2{\cal{L}}_0(\tra{E},\mu)\ln \left(\frac{\mu^2}{r^2 \nu^2}\right)+4{\cal{L}}_1(\tra{E},\mu)+\frac{\pi^2}{12}\delta(\tra{E}) \rbc,
\end{equation}
where compared with Eq.(\ref{eq:CS}) we have replaced $C_q$ with the general $C_B = (\bmat{T}_B)^2$, and similarly for the beam in the $n_{\overbar{B}}$ direction. The anomalous dimensions $\gamma_{\mu}^{sn,B}$ and $\gamma_{\nu}^{sn,B}$ are given in Eq.(\ref{eq:AD1}). We also note that the anomalous dimensions satisfy the following consistency relations,
\begin{equation}
  \label{eq:cons}
  \bmat{\Gamma}_{\mu}^{ss}(\tra{E})+\frac{1}{2}(\gamma_{\mu}^{sn,B}+\gamma_{\mu}^{sn,\overbar{B}})\delta(\tra{E})+\frac{1}{2}(\gamma_{\mu}^{sn,1}(\tra{E})+\gamma_{\mu}^{sn,2}(\tra{E})) = \bmat{\Gamma}^{\text{unmeas}}(\tra{E})\;\;, 
\end{equation}
and
\begin{equation}
  \gamma_{\nu}^{ss}(\tra{E})+\gamma_{\nu}^{sn,B}(\tra{E})+\gamma_{\nu}^{sn,\overbar{B}}(\tra{E})=0,
  \end{equation}
where $\bmat{\Gamma}^{\text{\unmeas}}$ is the anomalous dimension of the unmeasured and unfactorized  soft function for di-jet events. These consistency relations hold for both jet-veto and transverse energy measurements. For jet-veto measurements $\bmat{\Gamma}^{\text{unmeas}} (\veb{p})$ is given in Eq.(5.18) of Ref.~\cite{Hornig:2016ahz}. The corresponding expressions for the global and collinear soft functions and their anomalous dimensions  appropriate for jet veto measurements can be trivially evaluated by performing the integration over the transverse energy using the relations in Eq.(\ref{eq:PtoC}). It is now a simple exercise to confirm that from the product of global soft, $S_{s}(\veb{p})$, and collinear soft functions, $S_{n,i}(\veb{p})$, evaluated at a common scale $\mu$,  we recover the result of Ref.~\cite{Hornig:2016ahz} for the di-jet soft function in Eq.(4.28), and confirm the consistency relations  in Eq.(\ref{eq:cons}).

\subsection{Soft function renormalization group evolution}
\label{sec:3.2}

\subsubsection{Evolution in rapidity space}
\label{sec:rev}
As mentioned earlier, the $n_J$-collinear soft function is free of rapidity divergences and thus does not evolve in rapidity space. Here we discuss the RRG evolution of the global soft and $n_B$-collinear soft functions.  For the case of transverse energy measurements the solution of the RGE is discussed in the previous section below Eq.(\ref{eq:cons1}). For jet-veto measurements the rapidity space anomalous dimensions, $\gamma_{\nu}^{F}$, of the global and collinear soft functions is defined through the following RG equation, 
\begin{equation}
  \label{eq:jvRG}
  \frac{d}{d \ln \nu} F(\veb{p},R^{\text{veto}},\mu,\nu) = \gamma^F_{\nu}(\veb{p},R^{\text{veto}},\mu) F (\veb{p},R^{\text{veto}},\mu,\nu),
\end{equation}
where $R^{\text{veto}}$ is the jet cone size used in the jet vetoing process. In our perturbative calculations at NLO there is at most one parton in the vetoing region, thus it will be forming a single soft jet for all values of $R^{\text{veto}} > 0$. This is reflected in our calculations by the fact that none of the perturbative calculable elements of the factorization theorem depends on parameter $R^{\text{veto}}$ at this order. The  anomalous dimensions can be written in the following general form (see Eqs.(15) and (16) of Ref.~\cite{Stewart:2013faa}),
\begin{equation}
   \gamma^F_{\nu}(\veb{p},R^{\text{veto}},\mu) = 2 \Gamma_{\nu}^F [\alpha_s] \ln \lp\frac{\veb{p}}{\mu}  \rp + \Delta \gamma_{\nu}^F[\alpha_s,R^{\text{veto}}],
  \end{equation}
where $\Gamma_{\nu}^F [\alpha_s]$ and $\Delta \gamma_{\nu}^F [\alpha_s,R^{\text{veto}}]$ are given in Table.~\ref{tb:res}. Then, the solution to the RG equation in Eq.(\ref{eq:jvRG}) is
\begin{equation}
  \label{eq:VS}
   F (\veb{p},R^{\text{veto}},\mu,\nu) = \mathcal{V}(\veb{p},R^{\text{veto}},\mu,\nu,\nu_F)  F (\veb{p},R^{\text{veto}},\mu,\nu_F),
\end{equation}
where
\begin{equation}
  \label{eq:V}
\mathcal{V}(\veb{p},R^{\text{veto}},\mu,\nu,\nu_F) = \exp \lp \kappa_F^{\text{veto}}[\alpha_s,R^{\text{veto}},\nu,\nu_F]  \rp \lp\frac{\mu}{\veb{p}} \rp^{-  \eta_F^{\text{veto}}[\alpha_s,\nu,\nu_F]},
\end{equation}
with
\begin{equation}
  \eta_F^{\text{veto}}[\alpha_s,\nu,\nu_F]= 2 \Gamma_{\nu}^F [\alpha_s]  \ln\lp\frac{\nu}{\nu_F} \rp\;,  
\end{equation}
and
\begin{equation}
  \kappa_F^{\text{veto}}[\alpha_s,R^{\text{veto}},\nu,\nu_F]=\Delta\gamma_{\nu}^F[\alpha_s,R^{\text{veto}}] \ln\lp\frac{\nu}{\nu_F} \rp.
  \end{equation}
It should be noted that for jet-veto measurements the canonical rapidity scales are $\nu_{ss} \sim \veb{p}$  for the global soft function, and  $\nu_{sn_{B}} \sim \veb{p}/r$ for the $n_B$-collinear soft function.


\subsubsection{Evolution in virtuality space}
\label{sec:vev}

We will first discuss the transverse energy measurement. The global and $n_B$-collinear soft functions have the same virtuality canonical scales, $\mu_{ss}\sim \mu_{sn_{B}} \sim \tra{E}$. Thus we will be considering the simultaneous evolution of  of the global and $n_B$-collinear soft functions and separately the evolution of the $n_J$-collinear soft function.

The combined global and $n_B$-collinear soft function, $\bmat{S}_{s\text{-}n}(\mu)$, is defined by the convolution of the global soft and both $n_B$-collinear soft beam functions evolved up to a common rapidity scale, $\nu$,
\begin{equation}
\bmat{S}_{s\text{-}n}(\mu) \equiv \bmat{S}_s(\mu,\nu) \otimes S_{n,B}(\mu,\nu) \otimes S_{n,\overbar{B}}(\mu,\nu)\, ,
\end{equation}
and satisfy the following renormalization group equation
\begin{equation}
  \label{eq:RGsnB}
  \frac{d}{d \ln (\mu)} \bmat{S}_{s\text{-}n}(\tra{E},\mu) = \bmat{\Gamma}_{\mu}^{s\text{-}n}(\mu) \otimes \bmat{S}_{s\text{-}n}(\tra{E},\mu)+\text{h.c.}\;.
  \end{equation}
The anomalous dimension $\bmat{\Gamma}_{\mu}^{s\text{-}n}$ is given by the sum of the corresponding  anomalous dimensions,
\begin{equation}
  \bmat{\Gamma}_{\mu}^{s\text{-}n}(\tra{E})=\bmat{\Gamma}_{\mu}^{ss}(\tra{E})+\frac{1}{2}(\gamma_{\mu}^{sn_B} +\gamma_{\mu}^{sn_{\overbar{B}}})\delta(\tra{E}) = -\frac{\alpha_s}{\pi} \bmat{M}'\delta(\tra{E})+\frac{1}{2}\gamma_{\mu}^{s\text{-}n} (\tra{E})\;,
\end{equation}
where
\begin{equation}
  \gamma_{\mu}^{s\text{-}n} (\tra{E}) \equiv  2 \frac{\alpha_s}{\pi} \lb -(C_1+C_2)\mathcal{L}_0 (\tra{E},\mu)+\lp \sum_{i} C_i\ln \left( \frac{\bar{\mu_i}}{m_i} \right)+(C_B+C_{\overbar{B}})\ln(r) \rp \delta(\tra{E}) \rb.
\end{equation}
The anomalous dimension, $\gamma_{\mu}^{F}$ for both the $n_{J}$-collinear soft function and  for the combined global and $n_{B}$-collinear soft functions can be written in the following general form 
\begin{equation}
 \gamma_{\mu}^{F}(\tra{E},\mu)=-2\Gamma_{F}[\alpha_s] \mathcal{L}_0(\tra{E},\mu \;\xi_{F} ) +\gamma_{F}[\alpha_s] \delta(\tra{E}),
\end{equation}
where $\xi_F$ is a scaleless parameter and $\Gamma_{F}[\alpha_s]$ and $\gamma_{F}[\alpha_s]$ are the cusp and non-cusp part of the anomalous dimension, respectively,  with expansions in the strong coupling as described in Eqs.(\ref{eq:G}) and (\ref{eq:g}). Then the solution to Eqs.(\ref{eq:RGsnJ}) and (\ref{eq:RGsnB}) is given by
\begin{equation}
  S_{n,J}(\tra{E},\mu) =  \mathcal{U}_{sn,J}(\mu,\mu_0) \otimes S_{n,J}(\tra{E},\mu_0) ,
\end{equation}
and
\begin{equation}
  \bmat{S}_{s\text{-}n}(\tra{E},\mu) = \mathcal{U}_{s\text{-}n}(\mu,\mu_0) \otimes \lb \bmat{\Pi}^{\dag} (\mu_0,\mu)\bmat{S}_{s\text{-}n}(\tra{E},\mu_0)  \bmat{\Pi}(\mu_0,\mu)\rb,
  \end{equation}
respectively. The matrix $\bmat{\Pi}(\mu_0,\mu)$ is given in Eq.(\ref{eq:Pi})\footnote{Note that for the soft evolution the initial and final scales in the arguments of the matrix $\bmat{\Pi}$ are reversed compared to the hard function.  This is due the difference in the sign of the color non-trivial part of the soft and hard anomalous dimensions.} and the evolution kernels $\mathcal{U}_F$ are similar to the evolution kernels  discussed in Ref.~\cite{Ellis:2010rwa} for the measured jet and soft function evolution and are given by
\begin{equation}
  \mathcal{U}_{F}(\tra{E},\mu,\mu_F) = \exp\lp\gamma_E\; \omega_F(\mu,\mu_F)+K_{F}(\mu,\mu_F)\rp \frac{(\xi_F \; \mu_F)^{\omega_{F}(\mu,\mu_F)}}{\Gamma(-\omega_F)} \lb \frac{1}{\tra{E}^{1+\omega_F(\mu, \mu_{F})}} \rb_{+},
\end{equation}
where $\omega_F(\mu,\mu_F)$ and $K_F(\mu,\mu_F)$ are given formally in all order in perturbation  theory by Eq.(\ref{eq:Kt}) and (\ref{eq:wt}).

For the case of jet veto measurements the renormalization group equation takes form
\begin{equation}
    \frac{d}{d \ln (\mu)} F(\veb{p},R^{\text{veto}},\mu) = \gamma_{\mu}^{F}(\veb{p},R^{\text{veto}},\mu)  F(\veb{p},R^{\text{veto}},\mu),
\end{equation}
where
\begin{equation}
\gamma_{\mu}^{F}(\veb{p},R^{\text{veto}},\mu) = 2 \Gamma_F [\alpha_S] \ln \lp \frac{\mu \; \xi_F}{ \veb{p}} \rp +\gamma_{F}[\alpha_s,R^{\text{veto}}].
\end{equation}
The form of the RG equation is identical to the one described in Eq.(\ref{eq:unmeasRG}) and therefore the solution is given by Eq.(\ref{eq:U}) for $m_F = \veb{p}/\xi_F$. The values of $\Gamma_{F} [\alpha_s]$ and $\gamma_{F}[\alpha_s,R^{\text{veto}}]$ at one loop are identical to the transverse energy measurements and independent of $R^{\text{veto}}$. Therefore for the NLL and NLL' soft function, all ingredients for calculating the evolution kernels are given in Table~\ref{tb:res}.

\begin{table}[h!]
  \begin{center}
\begin{tabu}{|c|c|c|c|c|c|c|c|}
\hline
Function               & $\Gamma_F^0$  & $\gamma_F^0$ & $\mu_F$       & $\xi_F$ & $\Gamma_{\nu}^{F} $ & $\Delta\gamma_{\nu}^{F}$ & $\nu_F$ \\\hline \hline
$S_{n,J}$               & $-4C_J$       & 0           & $\tra{E} R$   & $1/R$   & --                 & --                      & --\\ \hline
 $\bmat{S}_{s}$             & --            & --          & $\tra{E}$     & 1       & $(C_B+C_{\overbar{B}})\alpha_s/\pi $ & $\mathcal{O}(\alpha_s^2)$ & $\tra{E}$ \\ \hline
$S_{n,B}$          & --            & --          & $\tra{E}$     & 1       & $-C_B\alpha_s/\pi $ & $\mathcal{O}(\alpha_s^2)$ & $\tra{E}/r$ \\ \hline
$\bmat{S}_{s\text{-}n}$  & $4 (C_1+C_2)$ & $ 4 \Delta \gamma_{s\text{-}n}$ & $\tra{E}$ & 1 & -- & -- & -- \\ \hline
\end{tabu}
  \end{center}
   \caption{Elements of soft functions evolution equations for transverse energy and jet-veto measurements. The parameters $m_F$ appearing in Eq.(\ref{eq:unmeasRG}) for the case of jet-veto measurements are given by $\veb{p}/ \xi_F$. In the last line  $ \Delta \gamma_{s\text{-}n} \equiv 2\sum_i C_i \ln(\bar{\mu}_i/m_i)+2(C_B+C_{\overbar{B}}) \ln(r)$.}
  \label{tb:res}
\end{table}


\subsection{Beam function}

The beam function describes the initial state radiation and can be expressed in terms of product of the quark field matrix elements and a delta function that imposes the measurements on the partons outside the beam cone,
\begin{equation}
  \mathcal{B}_{q/p}(\tra{E},x_B,\mu)=\sum_{X} \delta(\tra{E}-\tra{E}^X)\lngl p_n(k) \bigv \bar{\chi}_n(0) \frac{\gamma^- }{2} \bigv X \rngl \lngl X \bigv \delta (p^- -\overbar{\mathcal{P}}_n) \chi_n(0) \bigv p_n(k) \rngl,
\end{equation}
where $p_n(k)$ is the collinear proton with momentum $ k^{\mu} = (0^+,k^-,\prp{0})$, and  $x_B$ is the fraction of the proton momentum carried by the quark field. The large component of the  quark field light-cone momenta  is set to $p^-$ through the operator delta function $\delta(p^- -\overbar{P}_n)$ and therefore, $x_B=p^-/k^-$. The transverse energy of the state $X$ is given by  $\tra{E}^X$ and is defined as follows,
\begin{equation}
\tra{E}^X = \sum_{i \in X} \vert \bmat{p}_{T}^i \vert \Theta(\vebc{\eta} - \eta^i).
\end{equation}
In the $p^- r \gg \Lambda_{QCD}$ limit the beam function can be written as a convolution between collinear parton distribution functions (PDFs) and perturbative calculable coefficients~\cite{Stewart:2009yx}, 
\begin{equation}
  \label{eq:beam}
  \mathcal{B}_{j/p}(\tra{E},x_B,\mu) = \sum_i \int_{x_B}^1 \frac{dx}{x} \mathcal{I}_{j/i} (\tra{E},x_B,\mu)f_{i/p}\left(\frac{x_{B}}{x},\mu\right) + \mathcal{O} \Big{(} \frac{\Lambda_{\text{QCD}}}{(p^- r)^2}  \Big{)},
\end{equation}
where $\mathcal{I}_{j/i}$ are the short distance coefficients calculable in perturbation theory. In Eq.(\ref{eq:beam}) we suppress the dependence of $\mathcal{I}_{j/i}$ and $\mathcal{B}_{j/p}$ on $p^-$ and $r$ for simplicity. We organize the calculation of the beam function in a similar way as the calculation of the jet function in the last section. We denote  the contribution to the partonic beam function from radiation within the beam cone as $B_{i/j}(x,p^-,r)$. The contribution to the beam function from emissions outside the beam is $\Delta B_{i/j}(\tra{E},x,p^-,r)$, and $\mathcal{B}_{j/p}(\tra{E},x_B,\mu)$ is the sum of $B_{i/j}(x,p^-,r)$ and $\Delta B_{i/j}(\tra{E},x,p^-,r)$.

We can evaluate $B_{i/j}$ using the results for transverse virtuality of the incoming parton (i.e., the parton that enters the hard process), $t$, from Ref.~\cite{Ritzmann:2014mka}. At one-loop there is only one parton  contributing to the initial state radiation (ISR)  and its transverse momentum $\tra{p}$ is completely constrained by $t$ and $x$~\cite{Ritzmann:2014mka}:
\begin{equation}
  t=\left(\frac{x}{1-x} \right) \vert \vebv{p} \vert ^2,
\end{equation}
therefore the constraint that the ISR parton is within the beam cone with $r=e^{-\vebc{\eta}}$ can be expressed in terms of the transverse virtuality as follows,
\begin{equation}
  t<\left(\frac{1-x}{x} \right) (p^- r)^2,
\end{equation}
where we assume that is a massless on-shell parton. We can then rewrite the partonic level beam function at NLO as
\begin{equation}
  B_{i/j}^{\text{b}}(x,p^-,r) = \delta_{i/j} \delta(1-x)+  B^{\text{b},(1)}_{i/j}(x,p^-,r) + \mathcal{O}(\alpha_s),
\end{equation}
where for calculating the unmeasured contribution (i.e., restricting the parton within the beam cone) we have
\begin{equation}
  \label{eq:Bqq}
  B^{\text{b},(1)}_{q/q}(x,p^-,r) = \int d\Phi_2^{\text{c,ISR}}(t,x')\sigma_{2}^{\text{c}}(-t/x',1/x') \delta(x-x') \Theta_{\text{in}},
\end{equation}
and
\begin{equation}
  \label{eq:Bqg}
  B^{\text{b},(1)}_{q/g}(x,p^-,r) = -\frac{1}{1-\epsilon}\frac{T_F}{C_F}\int d\Phi_2^{\text{c,ISR}}(t,x')\sigma_{2}^{\text{c}}(-t/x',(x'-1)/x') \delta(x-x') \Theta_{\text{in}},
\end{equation}
where
\begin{equation}
\Theta_{\text{in}}=\Theta\Big{(}\frac{1-x}{x}  (p^- r)^2-t \Big{)},
\end{equation}
and $d\Phi_2^{\text{c,ISR}}(t,x)$ is the two-particle collinear phase space~\cite{Ritzmann:2014mka} and $\sigma_{2}^{\text{c}}$ the squared matrix element in $\overline{\text{MS}}$~\cite{Ritzmann:2014mka,Giele:1991vf}:
\begin{align}
  d\Phi_2^{\text{c,ISR}}(t,x) &= \frac{ [(1-x)t/x]^{-\epsilon}}{(4 \pi)^{2-\epsilon} \Gamma(1-\epsilon)}dxdt\;, & \sigma_{2}^{c}(s,x) & = \left(\frac{e^{\gamma_{E}}\mu^2}{4\pi}\right)^{\epsilon} \frac{2g^2}{s}P^{\text{b}}_{qq}(x).
\end{align}
where $P_{ij}^{\text{b}}$ are the bare QCD splitting kernels given in Ref.~\cite{ALTARELLI1977298,PhysRevD.9.980}. Details of the calculation are shown in Appendix~\ref{app:B} and here we simply summarize the results:
\begin{equation}
  \label{eq:finBqq}
  B_{q/q}^{\text{b},(1)}(x,p^-,r) = \frac{\alpha_s C_F}{2\pi}\lbc \frac{1}{\epsilon^2}\delta(1-x) + \frac{1}{\epsilon}\lb2 \delta(1-x)\ln\left(\frac{\mu}{p^- r} \right) -  \overbar{P}_{qq}(x) \rb \rbc +  \mathcal{I}^{(1)}_{q/q}(x,p^-,r),
\end{equation}
and
\begin{equation}
  \label{eq:finBqg}
  B_{q/g}^{\text{b},(1)}(x,p^-,r) = -\frac{\alpha_s T_F}{\pi} \frac{1}{2\epsilon} P_{qg}(x)  +  \mathcal{I}^{(1)}_{q/g}(x,p^-,r),
\end{equation}
where $\mathcal{I}_{i/j}^{(1)}$ are given in Eqs.(\ref{eq:I1qq}) and (\ref{eq:I1qg}) and the index b denotes that those are the ``bare'' quantities. The coefficients $\overbar{P}_{ij}$ can be expressed in terms of the standard QCD splitting kernels as follows,
\begin{align}
&\overbar{P}_{qq}(z)=P_{qq}(z)-\frac{3}{2}\delta(1-z)=(1+z^2){\cal{L}}_0(1-z), &
  &\overbar{P}_{qg}(z)=P_{qg}(z)=z^2+(1-z)^2,
\end{align}
We note that the $P_{ij}/\epsilon$ terms are interpreted as IR divergences that will cancel in the matching. The rest of the $\epsilon$ poles are UV divergences that need to be subtracted through renormalization. 

For the calculation of the contribution to the beam function from radiation outside the beam region, where we perform the measurement of the transverse energy, we need to consider the zero-bin subtractions which are not scaleless in this region. Both the zero-bin and collinear contributions suffer from divergences in the simultaneous limit $\tra{E} \to0$ and $x\to1$. We regulate these divergences using rapidity and dimensional regulators and we find as expected all divergences cancel in the final result. For transverse energy measurement at NLO we have,
\begin{align}
  \Delta B^{(1)}_{q/q}(\tra{E},x,p^-,r)&= \widetilde{\Delta B}_{q/q}^{(1)} - \Delta B^{\text{ z-bin},(1)}_{q/q} \nn \\ &= \frac{\alpha_s C_F}{\pi \tra{E}}  \lbc (1+x^2)\lb \frac{\Theta(x-x_0)}{1-x} \rb_+  
  +2\ln(x_0) \delta(1-x)\rbc ,
\end{align}
and
\begin{equation}
   \Delta B^{(1)}_{q/g}(\tra{E},x,p^-,r)= \widetilde{\Delta B}_{q/g}^{(1)} = \frac{\alpha_s T_F}{\pi} \frac{1}{\tra{E}} P_{qg}(x)\Theta(x-x_{0}),
\end{equation}
where $x_0=(1+\tra{E}/(p^- r))^{-1}$ and the explicit forms and the calculations of $\widetilde{\Delta B}_{i/j}^{(1)}$ and $\Delta B^{\text{ z-bin},(1)}_{i/j}$ are given Appendix~\ref{app:B}.

Since at one-loop there is only one parton contributing to ISR we can evaluate the beam functions for jet-veto measurements by integrating over the transverse energy before expanding in $\epsilon$ and $\eta$. Like for the case of soft function,  the beam function is independent of the jet size parameter $R^{\text{veto}}$ at this order. Performing the integrations and expanding first in $\eta$ and then in $\epsilon$ we get (for details of the calculation see Appendix~\ref{app:B}),
\begin{multline}
  \Delta B^{(1)}_{q/q,\text{veto}}(\veb{p},x,p^-,r)=\frac{\alpha_s C_F}{\pi} \lbc (1+x^2) \Big{(} \lb\frac{\Theta(x-x_0^{\text{vet.}})}{1-x}   \rb_+ \ln \left( \frac{\veb{p}x}{p^{-}r} \right)\\ - \lb \frac{\ln(1-x)}{1-x}\Theta(x-x_0^{\text{vet.}}) \rb_+  \Big{)}
  -\ln^2 (x_0^{\text{vet.}})\delta(1-x)
  \rbc,
\end{multline}
and
\begin{equation}
   \Delta B^{(1)}_{q/g,\text{veto}}(\veb{p},x,p^-,r) = \frac{\alpha_s T_F}{\pi} \lb \ln \left( \frac{\veb{p} x}{p^{-}r} \right)  - \ln(1-x)\rb P_{qg}(x) \Theta(x-x_0^{\text{vet.}}),
  \end{equation}
where for the case of jet-veto measurement  $x_0^{\text{vet.}}=(1+\veb{p}/(p^- r))^{-1}$.
The renormalized beam function is  defined through the following equation
\begin{equation}
  \mathcal{B}_{i/P}^{\text{b}}(\tra{E},x) = Z_{i}^{\mathcal{B}}(\mu)  \mathcal{B}_{i/P}(\tra{E},x,\mu),
\end{equation}
and satisfies the following RG equation
\begin{equation}
  \label{eq:beamRG}
\frac{d}{d \ln \mu} \mathcal{B}_{i/P}(\tra{E},x, \mu) = \gamma_{\mu}^{\mathcal{B},i}(\mu) \mathcal{B}_{i/P}(\tra{E},x,\mu),
  \end{equation}
where the index $i$ is not summed over and the renormalization function, $Z_{i}^{\mathcal{B}}$, and the anomalous dimension, $\gamma_{\mu}^{\mathcal{B},i}$, do not depend on the variable  $x$ or the transverse energy $\tra{E}$. Therefore in the $\overline{\text{MS}}$ scheme the complete matching coefficients $\mathcal{I}_{i/j}$ for the quark beam function in Eq.(\ref{eq:beam}) are :
\begin{equation}
  \mathcal{I}_{i/j} (\tra{E},x,\mu) = \lb \delta_{ij} \delta(1-x)+ \mathcal{I}^{(1)}_{i/j}(x,p^-,r)\rb \delta(\tra{E})+ \Delta B^{(1)}_{i/j}(\tra{E},x,p^-,r)+\mathcal{O}(\alpha_s^2),
\end{equation}
for transverse energy measurements and,
\begin{equation}
 \mathcal{I}_{i/j} (\veb{p},x,\mu) = \delta_{ij}\delta(1-x)+ \mathcal{I}^{(1)}_{i/j}(x,p^-,r)+ \Delta B^{(1)}_{i/j,\text{veto}}(\veb{p},x,p^-,r)+\mathcal{O}(\alpha_s^2),
\end{equation}
for jet-veto measurments. Based on  Eqs.(\ref{eq:beam}), (\ref{eq:I1qq}), and (\ref{eq:I1qg}) we find,
\begin{equation}
  \label{eq:bgamma}
  \gamma_{\mu}^{\mathcal{B},q}(p^-, r, \mu) = \frac{\alpha_s C_F}{\pi} \lb 2\ln \lp \frac{\mu}{p^- r} \rp + \bar{\gamma}_q\rb,
\end{equation}
where $\bar{\gamma}_q = 3/2$. We note that the anomalous dimension in the above equation is identical with the jet anomalous dimension after the replacement $p^- r \to \omega \tan(R/2)$ (see Eq.(6.26) of Ref.~\cite{Ellis:2010rwa}). This relation between beam and jet function is discussed in Appendix~\ref{app:B} below Eq.(\ref{eq:IJr}).  Therefore the unmeasured beam function is evolved as the unmeasured jet function with $\mu_\mathcal{B}=m_{\mathcal{B}}=p^- r$ and the solution of Eq.(\ref{eq:beamRG})  is given by Eq.(\ref{eq:U}) with the parameters given in Table~\ref{tb:appscales}.

\begin{figure}[t!]
   \centerline{\includegraphics[width = \textwidth]{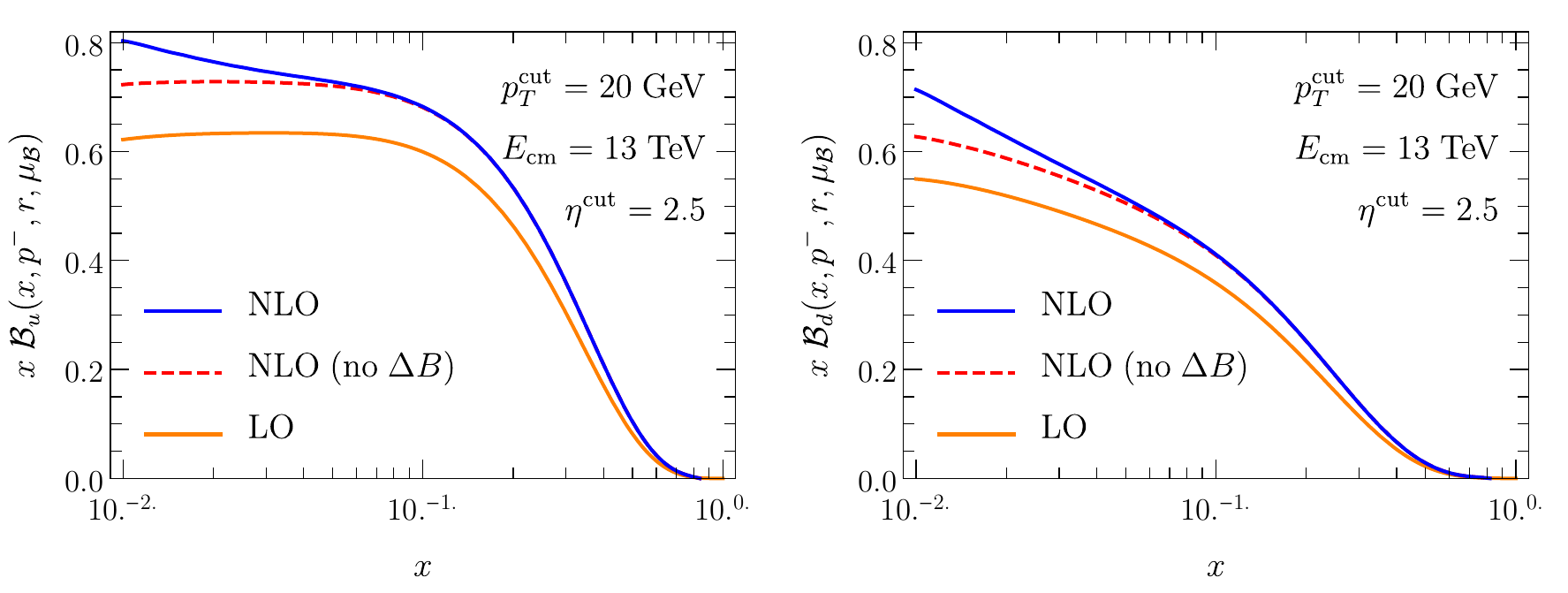}}
\caption{The LO (orange solid lines) and NLO (blue solid lines) beam function for the u-quark (left) and d-quark (right). For comparison we included the beam function at NLO without the corrections from out out of beam radiation, $\Delta B_{i/j}$,  (red dashed lines). All curves are evaluated at the beam scale $\mu_{\mathcal{B}}= x E_{\text{cm}}$. }
 \label{fig:beam}
 \end{figure}

In Fig.~\ref{fig:beam} we plot the LO and NLO beam functions for the up-quark (left) and down-quark (right). For the PDFs we use the CT10nlo data set extracted through the LHAPDF6 C++ library (see Ref.~\cite{Buckley:2014ana}).  Note that the LO beam function is simply the PDF evaluated at the scale $\mu=\mu_{\mathcal{B}}$. 
We choose $E_{\rm cm}=13$ TeV, $p_T^{\rm cut}=20$ GeV, and $\eta^{\rm cut}=2.5$. For comparison we also included the calculation of the beam function ignoring the NLO corrections from the out of beam radiation, $\Delta B_{i/j}$. Comparing the three curves we see that these corrections become important and  of the same size as the contribution from $\mathcal{I}_{i/j}^{(1)}$ for small values of $x \approx  \veb{p}/(E_{\text{cm}} r)$, corresponding to region II.e. In this region, fixed order QCD contributions are important. On the other hand, for higher values of $x$, region (II), for which $\veb{p} \ll p^{-} r$ they can be considered negligible .  This behavior is similar to what we found for the jet function corrections along the thrust axis in the electron-positron annihilation section (see Fig.~\ref{fig:outcorr}).

\section{Applications}
\label{sec:app}
We apply this analysis to the study of  di-jet cross sections in proton-proton collisions for measured or unmeasured jets with jet-veto measurements. This process was investigated within the framework of SCET in Ref.~\cite{Hornig:2016ahz} for $r \sim1$. In our analysis we take the small $r$ limit and we resum potentially large global logarithms of $r$, which will help reduce the theoretical uncertainty in the differential cross section. We also include corrections from contributions of ISR to the transverse energy and jet-veto measurement.

In this section we aim to show how using the formalism described in the last two sections can help us to reduce the theoretical uncertainty from scale variation in jet production cross sections. Thus, though gluon fusion processes ($gg \to gg$ and $gg \to q\bar{q}$) dominate  the di-jet cross section, here we consider the simpler example, $qq' \to qq'$. The complete calculation, which involves summing over all partonic channels, is beyond the scope of this work. 

The observable we are considering is the boost invariant version of angularities defined in Ref.~\cite{Hornig:2016ahz},
\begin{align}
\tau_a &\equiv \frac{1}{p_T} \sum_{i \in \text{jet}} p_T^i (\Delta\mathcal{R}_{iJ} )^{2-a}\;,&\;\;\text{with}\;\;\; \Delta\mathcal{R}_{iJ} &= \sqrt{(\Delta \eta_{iJ})^2 + (\Delta \phi_{iJ})^2},
\end{align}
where $a<1$ is the parameter the controls the wide angle radiation, $p_T$ is the transverse momentum of the jets,  $\Delta \eta_{iJ}$, and $\Delta \phi_{iJ}$ are the rapidity and azimuthal angle differences between the particle, $i$, and the jet, $J$, measured with respect to the beam axis. As an example we consider the case of one measured and one unmeasured jet for which the factorization theorem in SCET for region (II.e) is given by
\begin{multline}
    \frac{d\sigma^{\text{(II.e)}}}{dy_1 dy_2 dp_{T} d\tau_a} \equiv d\sigma (\veb{p},\tau_a) = \mathcal{N} \mathcal{B}_{q/P}(\veb{p},x_1,\mu) \mathcal{B}_{q'/P}(\veb{p},x_2,\mu)\\
      \Tr [\bmat{H}^{qq' \to qq'}(\mu)\bmat{S}(\veb{p},\tau_a,\mu)]\otimes_{\tau} J_1(\tau_a,\mu) J_2(\mu).
\end{multline}

The beam and soft functions appearing in the factorization theorem above are discussed in the previous section. The hard function $\bmat{H}$ is evaluated up to NLO in Refs.\cite{Kelley:2010fn}. The NLO expressions for the  measured and unmeasured jet functions, $J_{i}(\mu)$ and $J_{i}(\tau_a,\mu)$, for cone and $k_T$-type jet algorithms are given for the case of electron-positron annihilation in Ref.\cite{Ellis:2010rwa} and are generalized for $pp$ collisions in Eqs.(4.1) and (4.8) of Ref.~\cite{Hornig:2016ahz}. In the calculation of the jet functions both partons are constrained to be within the jet cone.  The contributions from the out-of-jet radiation are power suppressed by powers of $\tra{E}/p_T$ for the transverse energy and $\veb{p}/p_T$ for the jet veto measurement compared to the leading contributions of the corresponding $n_J$-collinear soft function in Eq.(\ref{eq:CSnJ}). This was discussed for the case of electron-positron annihilation in Ref.~\cite{Ellis:2010rwa} and is demonstrated for the case of $pp$ collisions in Appendix~\ref{app:C}.

All elements of the factorization theorem need to be evaluated at  the common scale $\mu$. The $\tau_a$ independent elements: beam, hard, and unmeasured jet function are evolved as described in Appendix~\ref{app:A.3}.  The evolution of  measured jet and soft functions is described in Sec. 6 of Ref.~\cite{Ellis:2010rwa}. Additionally the universal soft function, $\bmat{S}_{\text{unmeas}}$, is further factorized in the global soft function, $\bmat{S}_s$, and the collinear soft functions, $S_{n,i}$.  The  evolution of the $n_J$-collinear soft function, $S_{n,J}$, and the combination of global and $n_B$-collinear soft function, $\bmat{S}_{s\text{-}n}$, is described in Section~\ref{sec:vev}.  The global soft and the $n_{B}$-collinear functions, in addition to the evolution in virtuality space, also evolve in rapidity space. The evolution in rapidity space for transverse energy measurements is described in Section~\ref{sec:ee} and for jet-veto measurements in Section~\ref{sec:rev}.

We now have all the ingredients for the construction of the  cross section up to NLL' accuracy. For the choice $\mu=\mu_{ss}=\veb{p}$ we have,
\begin{multline}
  d\sigma^{\text{(II.e),NLL'}} (\veb{p},\tau_a) = \mathcal{N} \; \mathcal{U}_{\text{unmeas}}(\mu_H,\mu_{ss},\mu_J,\mu_J^{\tau},\mu^{}_{\mathcal{B}},\mu_{\overbar{\mathcal{B}}}) J_{q}(\mu_J)  \mathcal{B}_{q'/P}(\veb{p},x_2,\mu_{\overbar{\mathcal{B}}}) \\
  \times  \mathcal{B}_{q/P}(\veb{p},x_1,\mu^{}_{\mathcal{B}})  \; \lb \mathcal{U}_{\text{meas}}(\tau_a,\mu_J^{\tau},\mu_S^{\tau}) \lp 1+f_{q'}^{J}(\tau_a,\omega_S,\mu_J^{\tau})+f_{q'}^S(\tau_a,\omega_S,\mu_S^{\tau})\rp \rb_+ \\
 \times   \Tr \lb\bmat{\Pi}(\mu_{}ss,\mu_H)\bmat{H}^{qq'\to qq'}(\mu_H)\bmat{\Pi}^{\dag}(\mu_{ss},\mu_H)\bmat{S}_{\text{unmeas}}(\veb{p},\mu_{ss})\rb , 
\end{multline}
where 
\begin{equation}
 \mathcal{U}_{\text{unmeas}}(\mu_H,\mu_{ss},\mu_J,\mu_J^{\tau},\mu^{}_{\mathcal{B}},\mu_{\overbar{\mathcal{B}}}) = \prod_{F=\mathcal{B},\overbar{\mathcal{B}},J_1,J_2,H}\mathcal{U}_{F}(\mu_{ss},\mu_F).
\end{equation}
The functions $f_{i}^J(\tau_a,\omega_S,\mu_{J}^{\tau})$ and $f^S_{i}(\tau_a,\omega_S,\mu_S^{\tau})$ are given in Eqs.(5.11) and (5.26) of Ref.~\cite{Hornig:2016ahz} with $\omega_S(\mu_J^{\tau},\mu_S^{\tau})$ given in Eq.(\ref{eq:w}) and the elements of the anomalous dimension ($\Gamma_S^{0}$ and $\gamma_S^{0}$) are given in Table 2 of Ref.~\cite{Hornig:2016ahz}. The evolution kernel, $\mathcal{U}_{\text{meas}}$, evolves the measured soft function from it’s canonical scale up to the scale of measured jet function. After this point, the combined measured jet and soft function evolve as a jet down to the soft scale $\mu_S$, thus:
\begin{equation}
  \mathcal{U}_{\text{meas}}(\tau_a,\mu,\mu_0)= \frac{e^{K_S(\mu,\mu_0)+\gamma_E \omega_S(\mu,\mu_0)}}{\Gamma(-\omega_S(\mu,\mu_0))}\lp \frac{\mu_0}{m_S} \rp^{\omega_S(\mu,\mu_0)}\lb\frac{1}{\tau_a^{1+\omega_S(\mu,\mu_0)}} \rb_+ .
\end{equation}

For the choice $\nu=\nu_{ss}$ the universal part of the soft function, $\bmat{S}_{\text{unmeas}}(\veb{p},\mu_{ss})$ is given by
\begin{multline}
  \bmat{S}_{\text{unmeas}}(\mu_{ss}) = \bmat{S}_{ss}(\mu_{ss},\nu_{ss}) \lp \mathcal{V}_{sn,B}(\nu_{ss},\nu_{sn,B})S_{n,B}(\mu_{ss},\nu_{sn,B}) \rp^2 \\ \times \lp  \mathcal{U}_{sn,J}(\mu_{ss},\mu_{sn,J}) S_{n,J}(\mu_{sn,J}) \rp^2 .
\end{multline}
where we evolve in rapidity space the $n_B$-collinear soft function from $\nu_{sn,B} = \veb{p}/r$  to $\nu_{ss} = \veb{p}$ using Eqs.(\ref{eq:VS}) and (\ref{eq:V}). We also evolved in virtuality the $n_{J}$-collinear soft function from $\mu_{sn,J}= \veb{p} R$ to $\mu_{ss}$.

\begin{figure}[t!]
    \centerline{\includegraphics[width =0.5 \textwidth]{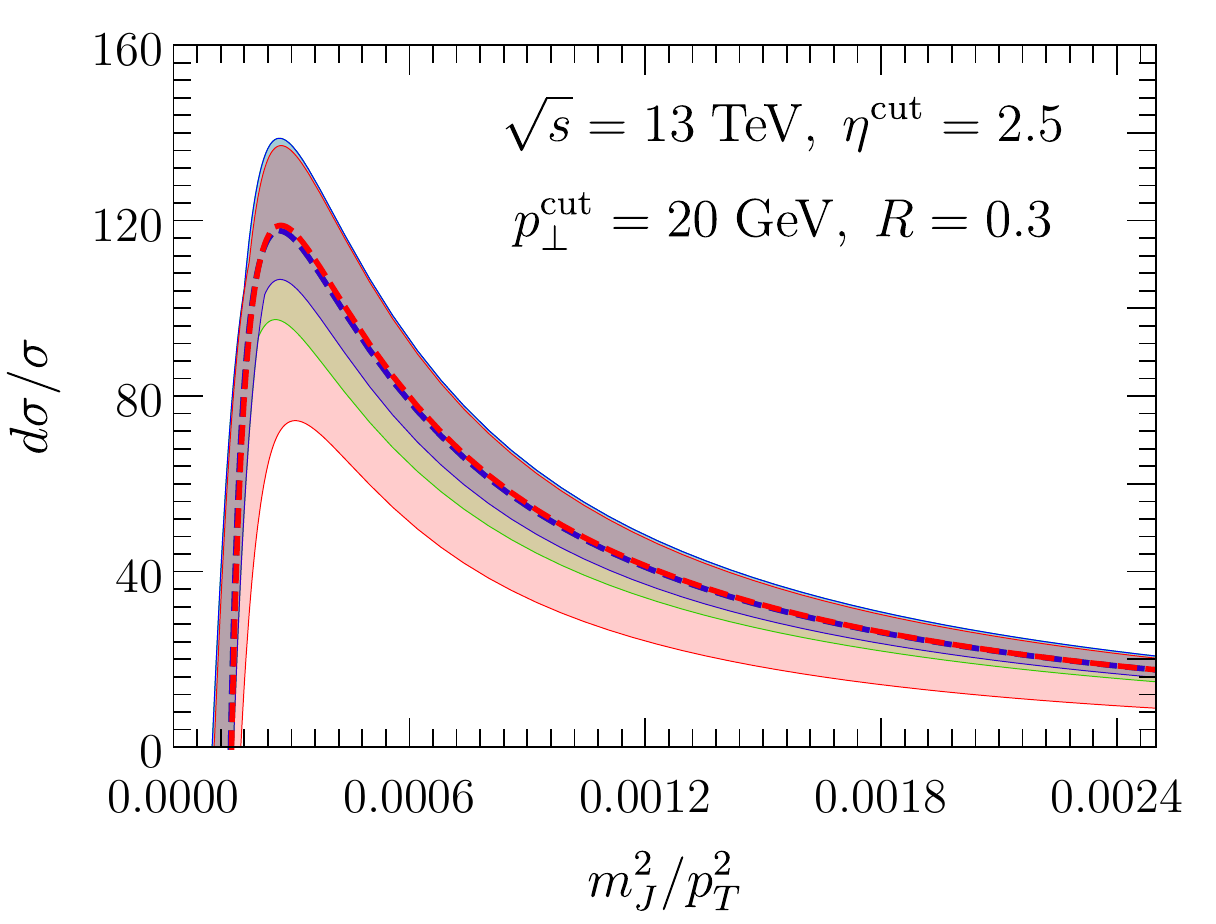}}
\caption{ The differential cross section as a function of $\tau_0 = m_J^2/\tra{p}^2+\mathcal{O}(\tau^2_0)$ for proton-proton collisions at $\sqrt{s} =13$ TeV for $\veb{p}=20$ GeV and $\tra{p}^{\text{jet}}= 500$ GeV. In this graph we used $\eta^{\text{cut}} = 2.5$ and the jet anti-$k_T$ algorithm with $R=0.3$. The red band corresponds to the complete unfactorized expression, the green band where only the jet-collinear modes are factorized, and blue band when both jet-collinear and beam-collinear modes are factorized in the unmeasured  soft function.  }
\label{fig:pptau2}
\end{figure}

We evaluate $d\sigma^{\text{(II.e)}}(\veb{p},\tau_a)$ as a function of $\tau_a$ for the following kinematic variables:
\begin{align}
   \veb{p}&=20 \text{ GeV}  &   p_T &=500\text{ GeV} &   E_{\text{cm}}&=13 \text{ TeV} & a&=0\nn \\ \vebc{\eta}& = 2.5 & R &=0.3 & \eta_{1} &= 1.0  & \eta_2 &= 1.4, 
\end{align}
and our results are  presented in Fig.~\ref{fig:pptau2}. For estimating the theoretical uncertainty we vary all canonical scales not associated with the jet shape measurement by 2 and 1/2. The jet measured and soft measured scales are varied within the profile function used in Ref.~\cite{Hornig:2016ahz}. In Fig.~\ref{fig:pptau2} the red band corresponds to the construction of the cross section using the  completely unfactorized soft function described in Section 4.3 of Ref.~\cite{Hornig:2016ahz}. The green band corresponds the global function in  which we factorized the jet-cone regions but not the beam-cone regions. This allows us to resum global logarithms of the jet-cone size parameter $R$ and thus leads to improved accuracy. Finally the blue band corresponds to the completely refactorized soft function where we factorize the beam-region as well, that way resuming  global logarithms of $r=e^{\vebc{-\eta}}$. We find that the refactorization of the soft function allows us to significantly improve the theoretical uncertainty.

\section{Conclusion}
\label{sec:conclusion}

In this paper we  consider the effect of rapidity cutoff in jet-veto and transverse energy measurements for exclusive  jet cross sections at LHC. We first demonstrate the effect of the vetoes in  electron-positron annihilations where the rapidity is measured with respect to the thrust axis and later extend this analysis to proton-proton collisions where the rapidity is  measured with respect to the beam axis. For the electron-positron anihilation analysis we find that two separate  factorization theorems are required to describe the transverse energy spectrum in the following  regions, region I: $\omega r \ll\prp{E} \ll \omega$,  and region II:  $\prp{E} \ll \omega r \ll \omega$  where $\omega = p^- \simeq 2 E_J$ is the large component of the jet light-cone  momentum and $r=\exp(\vebc{-\eta})$, where $\vebc{\eta}$ denotes the rapidity cutoff where veto is applied.

We find that for large transverse energy, i.e., region I, the cross section is insensitive to the exact value of $r$, as long as it satisfies the hierarchy describing this region. We show using \textsc{Pythia} simulations that for $ \prp{E}> 2 \omega r$ and for increasing transverse energy, the cross section $d\sigma(r)$ asymptotically reaches the limit $d\sigma(r \to 0)$ (see Fig.~\ref{fig:con}). This suggests that the factorization theorem appropriate for describing region I is independent of $r$ and involves the inclusive soft and jet functions for which no rapidity cutoff is implemented. The factorization of the cross section in this region was derived within the framework of SCET$_{\text{II}}$ in Ref.~\cite{Chiu:2012ir}.

In contrast, for region II the cross section is sensitive to the rapidity cutoff and therefore needs to be considered in the corresponding calculation. We propose a factorization theorem in SCET$_{\text{I}}$ which involves the unmeasured jet function calculated in Ref.~\cite{Ellis:2010rwa} and the refactorized soft function. The refactorization  of the soft function is necessary for resummation of global logarithms of $r$ that can be important in this region. Thus we employ the formalism introduced in Ref.~\cite{Chien:2015cka} and  separate contribution of global-soft modes  and soft-collinear modes. We find that the our analytic calculation at next-to-leading logarithmic prime  (NLL') accuracy agrees with the \textsc{Pythia} simulation within the theoretical uncertainty. Also the refactorization of the soft function helps reducing the theoretical uncertainty (see Fig.~\ref{fig:dig}). 

 Additionally, we consider corrections to the jet function from out-of-jet radiation when $\prp{E} \lesssim \omega r $ within the factorization theorem used for region II. We showed that including these corrections to the calculation of the NLL' cross-section  greatly improves the agreement of the  analytic results with the simulation data for $\prp{E} \lesssim \omega r $ (see Fig.~\ref{fig:outcorr}).

 In direct analogy from the electron-positron annihilation we extend our analysis to proton-proton collisions. We identify the two regions, region I  $p^- r \ll \tra{E} \ll p^-$,  and region II:  $\tra{E} \ll p^- r \ll p^-$  where the transverse energy, $\tra{E}$ is measured with respect to the beam axis and $p^- = x_B E_{\text{cm}}$ is the large light-cone component of the incoming parton in the hard process. Focusing on region II we use the formalism developed in Ref.~\cite{Chien:2015cka} to refactorize the soft function for exclusive jet production in proton-proton collisions. Unlike Ref.~\cite{Hornig:2016ahz}, here we consider the soft-collinear modes and functions along the beam direction. This allows us to resum  for the first time global logarithms of $\exp(-\vebc{\eta})$ to all orders in perturbation theory. The refactorized result involves the global soft function which is insensitive to the jet and beam cone boundaries,  and the beam and jet collinear soft functions  that take contributions from soft-collinear modes which can resolve the corresponding cone boundaries and therefore depend on the cone size parameters. In this work we study the cases of transverse energy and jet-veto measurements and  give the ingredients for constructing the 0,1, and 2-jet refactorized soft functions. As an example we study the di-jet cross section for the partonic channel $qq' \to qq'$ and we demonstrate that the refactorization of the soft function is necessary at NLL' accuracy  for keeping the theoretical uncertainty under control. 

 Furthermore we calculated for first time the perturbative ingredients for constructing the unmeasured quark beam function and we consider corrections from out-of-beam radiation. These corrections to the beam function allow us to extend the applicability of the factorization theorem for $\tra{E} \lesssim x_B E_{\text{cm}} e^{-\vebc{\eta}}$. We discussed the relation of our results to the fragmenting jet functions and the corresponding anomalous dimension.

 As an extension of this work we aim to complete the calculation of the di-jet cross section, including all the partonic channels. Furthermore, our analysis can be used to study the effect of underlying event activity in measurements of global observables such as the transverse energy within specific rapidity regions. An ideal process for such a measurement is isolated Drell-Yan at a large invariant mass compared to the typical soft scale of the underlying event.   

\acknowledgments
AH and DK are supported by the DOE Office of Science under Contract DE-AC52-06NA25396 and the Early Career Research Program (C. Lee, PI), and through the LANL/LDRD Program.
TM and YM are supported in part by the Director, Office of Science, Office of Nuclear Physics, of the U.S. Department of Energy under grant numbers DE-FG02-05ER41368.

\appendix

\section{Fixed order results and unmeasured evolution }
\label{app:A}

\subsection{Fixed order results for region I}
\label{app:A1}
Though in the factorization theorem in Eq.(\ref{eq:SCET2}) the jet function depends on the transverse momentum of the mother parton for evaluating the NLO cross section we only need the $\mathcal{O}(\alpha_s)$ terms when this transverse momentum is vanishing. The reason is because at this order the $\mathcal{O}(\alpha_s)$ terms of the jet function contribute only thought the LO soft function which is proportional to $\delta(\vebv{p}^2)$. Following similar arguments as in Section~\ref{sec:NLL(I)}, and the results in Eqs.(6.42) and (6.47) of Ref.~\cite{Chiu:2012ir} we get, 
\begin{multline}
  S(E_s,\vebv{p}^2,\vebv{q}^2)= \delta(E_s)\delta(\vebv{p}^2)\delta(\vebv{q}^2) + \frac{\alpha_s C_F}{2\pi}\lbc 2\; {\cal{L}}_0(E_s,\mu) \ln \left(\frac{\nu^2}{\mu^2}\right) -4\; {\cal{L}}_1 (E_s,\mu) -\frac{\pi^2}{12} \delta(E_s) \rbc \\
   \times\lp \delta ( E_s^2-\vebv{p}^2)\delta(\vebv{q}^2)+ \lbrack q \leftrightarrow p \rbrack \rp,
\end{multline}
and
\begin{equation}
  J_{(q)}(E_n,0)=\delta(E_n- p_{\perp})+ \frac{\alpha_s C_F}{2\pi}\lbc \frac{1}{2} \delta ( E_n) -\lb 3+2 \ln \left(\frac{\nu^2}{\omega^2}\right) \rb {\cal{L}}_0 (E_n,\mu) \rbc,
\end{equation}
which suggest the following canonical choices for the jet and soft scales,
\begin{align}
  \mu_s = \mu_J&=\prp{E},& \nu_S=\prp{E},& \; &\nu_J=\omega. 
  \end{align}
The NLO cross section for $\prp{E} \ll \omega $ can be constructed using the factorization theorem in SCET$_{\text{II}}$ given by Eq.(\ref{eq:SCET1})
\begin{multline}
  \frac{d\sigma}{d\prp{E}}^{\text{(I),NLO}}= \sigma_0 H_2  \lb \delta(\prp{E})-\frac{\alpha_s C_F}{2 \pi} \lbc  8\; {\cal{L}}_1 (\prp{E},2\mu) +\lb6+4\ln \lp\frac{\mu^2}{\omega^2}\rp\rb {\cal{L}}_0 (\prp{E},2\mu) \\ -\lb1-\frac{\pi^2}{6} \rb \delta(\prp{E}) \rbc \rb.
\end{multline}

We note that this result is independent of the rapidity scale, $\nu$, as expected from the cancellation of rapidity divergences between soft and jet function. For the integrated cross section defined in Eq.(\ref{eq:cum}) we use 
\begin{align}
\label{eq:PtoC}
&\int_0^{\Lambda} dE \; {\cal{L}}_0 \left(E,\mu \right) = \ln \left( \frac{\Lambda}{ \mu}\right)\;, 
&\int_0^{\Lambda} dE \; {\cal{L}}_1 \left(E,\mu \right) = \frac{1}{2}\ln^2 \left( \frac{\Lambda}{ \mu}\right)\;,  
\end{align}
to obtain,
\begin{equation}
d\sigma^{\text{(I),NLO}}(\vebp{p})= \sigma_0 H_2  \lb 1+\frac{\alpha_s C_F}{2 \pi} \lbc-4\ln^2 \lp \frac{\vebp{p}}{2 \omega}\rp+\ln^2 \lp \frac{\mu^2}{\omega^2}\rp -6\ln \lp \frac{\vebp{p}}{2 \mu}\rp +1-\frac{\pi^2}{6}  \rbc \rb.
\end{equation}
Using the hard function at NLO which is~\cite{Bauer:2003di,Manohar:2003vb}
\begin{equation}
H_2(\omega,\mu)=1-\frac{\alpha_s C_F}{2\pi} \lbc 8-\frac{7 \pi^2}{6}+\ln^2 \left(\frac{\mu^2}{\omega^2} \right)+3\ln \left(\frac{\mu^2}{\omega^2} \right) \rbc,
\end{equation}
we get 
\begin{equation}
\frac{1}{\sigma_0}d \sigma^{\text{(I),NLO}}(\vebp{p})=1+\frac{\alpha_s C_F}{2\pi} \lbc -7+\pi^2-6\ln \left(\frac{\vebp{p}}{2 \omega} \right)-4\ln^2 \left(\frac{\vebp{p}}{2 \omega} \right) \rbc.
\end{equation}
which agrees with the full theory result for $\vebp{p} > 2 \omega r$ and $\vebp{p} \ll \omega$.

\subsection{Fixed order jet function for region II}

Using the results of Ref.~\cite{Ellis:2010rwa} for the diagrams that contribute to the calculation of the jet function at 1-loop, the jet function for the transverse energy measurement with rapidity cutoff and can be written in the following form:
\begin{equation}
  \tilde{\mathcal{J}}_{q}^{\text{b}}(\prp{E},\omega,r)= 2\frac{\alpha_s C_F}{2 \pi} \frac{(e^{ \gamma_E} \mu^{2})^{\epsilon}}{\Gamma(1-\epsilon)} \Big{(}\frac{\nu}{\omega} \Big{)}^{\eta}  \int dx d\prp{k} \; \frac{1}{\prp{k}^{1+2 \epsilon}}  \lb 2\frac{1-x}{x^{1+\eta}}+(1-\epsilon)x  \rb \prp{\Theta} , 
\end{equation}
where $x\equiv k^-/\omega$ is the portion of the original parton energy carried by the gluon and we also use the rapidity regulator since the naive result for our measurement contains rapidity divergences though they cancel when adding the zero-bin subtraction.  Also we define
\begin{multline}
  \prp{\Theta}=  \Theta \Big{(} x \omega r - \prp{k} \Big{)} \Theta \Big{(} (1-x)\omega r - \prp{k}  \Big{)}  \delta(\prp{E})
  + \lb \Theta \Big{(}\prp{k} - x \omega r \Big{)} \Theta\Big{(} (1-x)\omega r - \prp{k}\Big{)} \\
  + \Theta\Big{(}  x \omega r - \prp{k}  \Big{)} \Theta\Big{(}\prp{k} - (1-x)\omega r\Big{)}  \rb \delta(\prp{E}-\prp{k}).
\end{multline}
The first term in $\prp{\Theta}$  is rapidity divergence free and corresponds to the case where both daughter pardons are emitted within the unmeasured region. The contribution from this term is calculated in Ref.~\cite{Ellis:2010rwa} and after renormalization is given by $J_{q}$ in Eq.(\ref{eq:jet}). The second and third terms correspond to the case where either the quark or gluon only is emitted inside the measured region, respectively. The divergences appearing in the second term are only rapidity divergences and the third term is finite. The case where both patrons are emitted within the unmeasured region is not included since this region of phase space contributes only for $\prp{E} > \omega r $. In the Region II the contributions of second and third term, which we will denote as $\widetilde{\Delta J}_{q}$, are power suppressed and can be ignored, thus the jet function reduces to the unmeasured jet function in Eq.(\ref{eq:jet}). On the other hand in Region II.e those terms become of $\mathcal{O}(1)$ and need to be included. Performing the integration we have
\begin{multline}
  \label{eq:navjetc}
 \widetilde{\Delta J}_{q}(\prp{E},\omega,r)=  \frac{\alpha_s C_F}{2\pi} \frac{1}{\prp{E}} \lbc -4 \lb\frac{1}{\eta}+\ln \Big{(} \frac{\nu}{\omega} \Big{)}-\ln \Big{(} \frac{\prp{E}}{\omega r} \Big{)} \rb\\-\Theta(\prp{E})\Theta(\omega r/2-\prp{E})\lb 6 \frac{\prp{E}}{\omega r}+ 4\ln \Big{(}1- \frac{\prp{E}}{\omega r} \Big{)} \rb \\+\Theta(  \prp{E} - \omega r/2)\Theta(\omega r-\prp{E}) \lb  6 \frac{\prp{E}}{\omega r}-6-8\ln \Big{(}\frac{\prp{E}}{\omega r} \Big{)}+ 4\ln \Big{(}1- \frac{\prp{E}}{\omega r} \Big{)}  \rb\rbc ,
  \end{multline}

Correspondingly the zero-bin subtractions can be constructed  by taking the leading contributions in the $x \to 0 $ limit
\begin{equation}
  \mathcal{J}_{q}^{\text{ z-bin},(1)}(\prp{E},\omega,r)= 2\frac{\alpha_s C_F}{2 \pi} \frac{(e^{ \gamma_E} \mu^{2})^{\epsilon}}{\Gamma(1-\epsilon)}   \int \frac{ dx d\prp{k}}{\prp{k}^{1+2 \epsilon}} \;\frac{2}{x^{1+\eta}} \prp{\Theta}^{\text{ z-bin}}  ,
\end{equation}
where
\begin{equation}
  \prp{\Theta}^{\text{ z-bin}}= \prp{\Theta} \Big{\vert}_{x \ll 1} \simeq \Theta\Big{(}x \omega r - \prp{k} \Big{)}  \delta(\prp{E})+\Theta\Big{(}\prp{k}- x \omega r \Big{)}  \delta(\prp{E}-\prp{k}).
\end{equation}
We find that for the zero-bin subtraction we only have two contributions. The first term in $\prp{\Theta}^{\text{ z-bin}}$ gives the contribution when the soft gluon is emitted within the jet cone (unmeasured region) and the second term corresponds to case where the soft gluon is emitted within measured region outside the jet cone. The first term reduces to a scaleless  integral which we ignore and the second term will contribute to $\Delta J_{q}^{(1)}$:
\begin{equation}
  \Delta J_{q}^{\text{ z-bin},(1)}(\prp{E},\omega,r)=  -\frac{\alpha_s C_F}{2\pi} \frac{4}{\prp{E}}  \lb\frac{1}{\eta}+\ln \Big{(} \frac{\nu}{\omega} \Big{)}-\ln \Big{(} \frac{\prp{E}}{\omega r} \Big{)} \rb.
\end{equation}
The final result of the total jet function we have
\begin{equation}
  \mathcal{J}_{q}^{\text{NLO}}(\prp{E},\omega,r) = J_{q}^{\text{NLO}}(\omega, r) \delta(\prp{E}) + \Delta J_{q}^{(1)}(\prp{E},\omega,r),
\end{equation}
where
\begin{multline} 
  \Delta J_{q}^{(1)} (E_{\perp},\omega,r)\;= \widetilde{\Delta J}_{q}^{(1)}(\prp{E},\omega,r)-\Delta J_{q}^{\text{ z-bin},(1)}(\prp{E},\omega,r)\\
   =-\frac{\alpha_s C_F}{2\pi} \lbc  \Theta(\prp{E})\Theta(\omega r/2-\prp{E}) \lb 6 \left(\frac{1}{r \omega} \right) + \frac{4}{E_{\perp}} \ln \left(1-\frac{E_{\perp}}{r \omega} \right) \rb  \\
- \Theta(  \prp{E} - \omega r/2)\Theta(\omega r-\prp{E}) \lb 6 \left(\frac{1}{r \omega} \right) - \frac{6}{E_{\perp}}+ \frac{4}{E_{\perp}} \ln \left(1-\frac{E_{\perp}}{r \omega} \right) - \frac{8}{E_{\perp}} \ln \left(\frac{E_{\perp}}{r \omega} \right) \rb \rbc.
\end{multline}


\subsection{Renormalization group evolution of jet and hard functions}
\label{app:A.3}

Unmeasured quantities such as the hard function in region I  and the jet functions in region II satisfy the following renormalization group equations:
\begin{equation}
  \label{eq:unmeasRG}
  \mu \frac{d}{d\mu}F(\mu)= \lb \Gamma_F[\alpha_S] \ln \left(\frac{\mu^2}{m^2_F} \right) + \gamma_F[\alpha_S]\rb F(\mu),
\end{equation}
where $\Gamma_F$ is the cusp part of the anomalous dimension which is proportional to the cusp anomalous dimension and has the following expansion in the strong coupling
\begin{equation}
  \label{eq:G}
\Gamma_F[\alpha_s] =  (\Gamma_F^0/\Gamma_{\text{cusp}}^0) \Gamma_{\text{cusp}} = (\Gamma_F^0/\Gamma_{\text{cusp}}^0) \sum_{n=0}^{\infty} \left(\frac{\alpha_s}{4 \pi} \right)^{1+n} \Gamma_{\text{cusp}}^n.
\end{equation}
Similarly the non-cusp part of the anomalous dimension, $\gamma_F$, has the following expansion 
\begin{equation}
   \label{eq:g}
\gamma_F[\alpha_s] =  \sum_{n=0}^{\infty} \left(\frac{\alpha_s}{4 \pi} \right)^{1+n} \gamma_{F}^n.
\end{equation}
The solution of Eq.(\ref{eq:unmeasRG}) is
\begin{align}
  \label{eq:U}
F(\mu)&= \mathcal{U}_F(\mu,\mu_0) F(\mu_0) \, , &\mathcal{U}_F(\mu,\mu_0)=\exp \left( K_F (\mu, \mu_0) \right) \left( \frac{\mu_0}{m_F} \right) ^{\omega_F(\mu, \mu_0)},
\end{align}
where formally to all orders in perturbation theory the exponents $K_F$ and $\omega_F$ are given by,
\begin{align}
  \label{eq:Kt}
K_F(\mu, \mu_0) &= 2 \int_{\alpha (\mu_0)}^{\alpha(\mu)} \frac{d \alpha}{\beta(\alpha)} \Gamma_F (\alpha) \int_{\alpha(\mu_0)}^{\alpha}
\frac{d \alpha'}{\beta(\alpha')} +\int_{\alpha (\mu_0)}^{\alpha(\mu)} \frac{d \alpha}{\beta(\alpha)} \gamma_F (\alpha) ,\\
\label{eq:wt}
\omega_F(\mu, \mu_0) &= 2 \int_{\alpha (\mu_0)}^{\alpha(\mu)} \frac{d \alpha}{\beta(\alpha)} \Gamma_F (\alpha).
\end{align}
For  NLL and NLL' accuracy, which we are considering in this work,
\begin{align}
    \label{eq:K}
K_F(\mu, \mu_0) &=-\frac{\gamma_F^0}{2 \beta_0} \ln r -\frac{2 \pi \Gamma_F^0}{(\beta_0)^2} \Big{\lbrack} \frac{r-1+r\ln r}{\alpha_s(\mu)}
+ \left( \frac{\Gamma^1_c}{\Gamma^0_c}-\frac{\beta_1}{\beta_0} \right) \frac{1-r+\ln r}{4 \pi}+\frac{\beta_1}{8 \pi \beta_0}
\ln^2 r  \Big{\rbrack}, \\
\label{eq:w}
\omega_F(\mu, \mu_0) &= - \frac{\Gamma_F^0}{ \beta_0} \Big{\lbrack} \ln r + \left( \frac{\Gamma^1_c}{\Gamma^0_c} -
\frac{\beta_1}{\beta_0}  \right) \frac{\alpha_s (\mu_0)}{4 \pi}(r-1)\Big{\rbrack},
\end{align}
where $r=\alpha(\mu)/\alpha(\mu_0)$ and $\beta_n$ are the coefficients of the QCD $\beta$-function,
\begin{equation}
\beta(\alpha_s) = \mu \frac{d \alpha_s}{d \mu}= -2 \alpha_s \sum_{n=0}^{\infty} \left( \frac{\alpha_s}{4 \pi} \right)^{1+n} \beta_n \; .
\end{equation}
\begin{table}[t!]
  \begin{center}
\begin{tabu}{|c|c|c|c|c|}
\hline
Function & $\Gamma_F^0$ & $\gamma_F^0$ & $m_F(e^+ e^-)$ & $m_F(pp)$\\
\hline \hline
$ H(\omega,\mu) $                       & $ - 4 \sum_{i} C_{i} $ & $- 4 \sum_i \bar{\gamma}_i$ & $\omega $ & $m_i$ \\ \hline
$ J_{n}^{(i)}(\omega, R,\mu) $ &  $ 4 C_{i} $  &  $ 4 \bar{\gamma}_i $ &  $ \omega \tan(R/2) $ & $ p_T R$\\ \hline
$ B_{i/P}(p^-, r,\mu) $ &  $ 4 C_{i} $  &  $ 4 \bar{\gamma}_i $ &  n.a.  & $ p^- r$\\ \hline
\end{tabu}
 \caption{Evolution table: $\bar{\gamma}_q = 3 C_F/2$ and $\bar{\gamma}_g=\beta_0/2$.}
  \label{tb:appscales}
\end{center}
\end{table}
For unmeasured functions the scale $m_F$ equals the canonical  scale of the perturbative  function $F$. For the  hard and jet functions for electron-positron annihilation the quantities $\Gamma_F^0$, $\gamma_F^0$, and $m_F$ are summarized in Table~\ref{tb:appscales}. For the di-jet hard function in hadronic collisions, the evolution of the hard function is complicated due to the non-trivial color structure. The evolved hard function is given in Section 5.1 of Ref.~\cite{Hornig:2016ahz}:
\begin{equation}
  \bmat{H}(\mu)= \mathcal{U}_H(\mu,\mu_0) \bmat{\Pi}(\mu,\mu_0) \bmat{H}(\mu_0) \bmat{\Pi}^{\dag}(\mu,\mu_0),
\end{equation}
where
\begin{equation}
  \label{eq:Pi}
   \bmat{\Pi}(\mu,\mu_0)=\exp \lbc \bmat{M}' \int_{\alpha_s(\mu_0)}^{\alpha_s(\mu)} \frac{d\alpha}{\beta[\alpha]}\Gamma_{\text{cusp}} \rbc,
\end{equation}
where $\bmat{M}'$ is given in Eq.(\ref{eq:Mp}). The kernel $\mathcal{U}_H$ can be constructed from Eqs.(\ref{eq:U}),(\ref{eq:K}), and (\ref{eq:w}) and Table~\ref{tb:appscales}.
In comparison to the results of Ref.~\cite{Kelley:2010fn} we omitted the $i\pi \bmat{T}$ term since, as discussed in Ref.~\cite{Hornig:2016ahz}, this term cancels in the RGE when we sum the hermitian conjugate term.



\section{Fixed order results for quark beam function }
\label{app:B}

In this section we give some more details of the calculation of beam function for measured transverse energy and jet-veto measurements outside the beam region. From Eq.(\ref{eq:Bqq}),
\begin{align}
  \label{eq:calcBqq}
  B_{q/q}^{\text{b},(1)}(x,p^-,r) &= \int d\Phi_2^{\text{c,ISR}}(t,x')\sigma_{2}^{\text{c}}(-t/x',1/x') \delta(x-x') \Theta_{\text{in}} \nn \\
  &=\frac{\alpha_s C_F}{2 \pi} \frac{(e^{\gamma_E} x \mu^2)^{\epsilon}}{\Gamma(1-\epsilon)} \lb \frac{1+x^2}{(1-x)^{1+\epsilon}} -\epsilon (1-x)^{1-\epsilon}  \rb \int^{(1-x)(p^- r)^2/x}_{0} \frac{dt}{t^{1+\epsilon}} \nn\\
  &=-\frac{\alpha_s C_F}{2 \pi} \frac{(e^{\gamma_E} x^2)^{\epsilon}}{\epsilon\Gamma(1-\epsilon)} \left(\frac{\mu}{p^- r} \right)^{2\epsilon}\lb \frac{1+x^2}{(1-x)^{1+2\epsilon}} -\epsilon (1-x)^{1-2\epsilon}  \rb.
\end{align}
where in the first line,
\begin{equation}
\Theta_{\text{in}}=\Theta\Big{(}\frac{1-x}{x}  (p^- r)^2 -t \Big{)}.
\end{equation}
Expanding Eq.(\ref{eq:calcBqq}) in $\epsilon$ we get the result in Eq.(\ref{eq:finBqq}) with
\begin{multline}
  \label{eq:I1qq}
  \mathcal{I}^{(1)}_{q/q}(x,p^-,r) = \frac{\alpha_s C_F}{\pi} \lbc \frac{1}{2}(1-x)+\delta(1-x) \lb \ln^2 \left(\frac{\mu}{p^-r} \right)-\frac{\pi^2}{24} \rb+(1+x^2)\mathcal{L}_1(1-x)\\ -\overbar{P}_{qq}(x) \ln \left( \frac{x\mu}{p^-r} \right)\rbc.
  \end{multline}
Similarly for the contribution from the gluon we have from Eq.(\ref{eq:Bqg})
\begin{align}
  B_{q/g}^{\text{b},(1)}(x,p^-,r) &= -\frac{1}{1-\epsilon}\frac{T_F}{C_F}\int d\Phi_2^{\text{c,ISR}}(t,x')\sigma_{2}^{\text{c}}(-t/x',(x'-1)/x') \delta(x-x') \Theta_{\text{in}} \nn \\
  &=\frac{\alpha_s T_F}{2\pi} (1+\epsilon) \mu^{2\epsilon} \left(\frac{x}{1-x} \right)^{\epsilon} \lb x^2 +(1-x)^2 -\epsilon \rb \int^{(1-x)(p^- r)^2/x}_{0} \frac{dt}{t^{1+\epsilon}} \nn\\
  &=-\frac{\alpha_s T_F}{2\pi}\left(\frac{1}{\epsilon}+1 \right)\left(\frac{x}{1-x} \right)^{2\epsilon} \left(\frac{\mu}{p^- r} \right)^{2\epsilon}\lb P_{qg}(x)-\epsilon  \rb.
\end{align}
Expanding the above equation in $\epsilon$ we get the result in Eq.(\ref{eq:finBqg}) with
\begin{equation}
   \label{eq:I1qg}
 \mathcal{I}^{(1)}_{q/g}(x,p^-,r) =\frac{\alpha_s T_F}{\pi} \lbc x(1-x)+P_{qg}(x) \lb \ln \left( \frac{p^- r}{x \mu}  \right) + \ln(1-x) \rb \rbc.
\end{equation}

Since the above results correspond to the case where the ISR parton is emitted within the beam region where no measurement is performed, these are universal for any measurement:
\begin{equation}
   B_{q/q}(e,x,p^-,r)=B_{q/q}(x,p^-,r) \delta(e),
\end{equation}
where $e$ is the measured observable. For jet-veto like measurements we simply have
\begin{equation}
   B_{q/q}(\vebc{e},x,p^-,r)=B_{q/q}(x,p^-,r). 
\end{equation}
For the case where the parton is emitted within the phase-space region where measurements are performed we need to calculate the contribution for each measurement independently. In the next two sections we give the details of the calculation for transverse energy and jet-veto measurements.

We note here that at this order the matching coefficients, $\mathcal{I}^{(1)}_{i/j}(x,p^-,r)$, are related to the matching coefficients, $\mathcal{J}^{(1)}_{j/i}(x,\omega \tan(R/2))$, of the unmeasured fragmenting jet function onto the collinear fragmentation function from Ref.\cite{Procura:2011aq} through the replacement $\omega \tan(R/2) \to p^- r/x$: 
\begin{equation}
  \label{eq:IJr}
  \mathcal{I}_{i/j}(x,p^-,r)\Big{\vert}_{0<x<1} = \mathcal{J}_{j/i}(x, p^- r/x) \Big{\vert}_{x>1/2}.
\end{equation}
This then implies the following relation between the beam and jet anomalous dimensions:
\begin{equation}
 \label{eq:jetb}
  \gamma_{\mu}^{B}(\mu_B) = \gamma_{\mu}^{J}(\mu_{J}\to\mu_{B}),
\end{equation}
where $\mu_B = p^- r$ and $\mu_J = \omega \tan(R/2)$. This is shown explicitly for the quark beam function at one loop in Eq.(\ref{eq:bgamma}). These relations in Eq.(\ref{eq:IJr}) can be easily checked at NLO for the cases $\mathcal{I}_{q/q}$ and $\mathcal{I}_{q/g}$ using the results of this section and we  believed that hold for  $\mathcal{I}_{g/g}$ and  $\mathcal{I}_{g/q}$ as well. The explicit calculations for the remaining two cases is left for a subsequent publication.

\subsection{Transverse energy measurement}

As already mentioned in Section~\ref{sec:pp} contributions to the beam function from emission of partons within the measured region of phase-space suffer from rapidity divergences that need to be regulated. Additionally the soft-bin subtractions do not give scaleless integrals and thus will also contribute to the calculation of the beam function. Furthermore, the soft-bin contributions themselves require rapidity regulator and as we will show the total results turns out be finite and independent of the rapidity regulator parameters. The correction to the beam function is 

\begin{multline}
 \label{eq:calcDBqq}
 \widetilde{\Delta B}_{q/q}^{(1)}(\tra{E},x,p^-,r) = \int d\Phi_2^{\text{c,ISR}}(t,x')\sigma_{2}^{\text{c}}(-t/x',1/x')\delta(x-x') \left(\frac{\nu}{(1-x)p^-} \right)^{\eta}\Theta_{\text{meas.}}   \\
 = \frac{\alpha_s C_F}{\pi} \frac{(e^{\gamma_E} \mu^2)^\epsilon}{\Gamma(1-\epsilon)}\Big{(}\frac{\nu}{p^-} \Big{)}^{\eta}  \lb \frac{1+x^2}{(1-x)^{1+\eta}} -\epsilon (1-x)^{1-\eta}  \rb  \frac{\Theta(x-x_0)}{\tra{E}^{1+2\epsilon}} ,
\end{multline}
where
\begin{align}
  \Theta_{\text{meas.}}&=(1-\Theta_{\text{in}})\delta\Big{(}\tra{E}-\lb\frac{(1-x)t}{x}\rb^{1/2} \Big{)}  & x_0 &= \lb 1+\frac{\tra{E}}{p^- r} \rb^{-1},
\end{align}
and we used
\begin{equation}
  \delta\Big{(}\tra{E}-\lb\frac{(1-x)t}{x}\rb^{1/2}\Big{)}= 2 \frac{x}{1-x} \tra{E} \delta \Big{(} t- \frac{x}{1-x}\tra{E}^2 \Big{)}.
  \end{equation}
For the soft-bin subtraction
\begin{align}
  \label{eq:calcDBqqS}
  \Delta B^{\text{ z-bin},(1)}_{q/q}(\tra{E},x,p^-,r) &= 4\left(\frac{e^{\gamma_{E}}\mu^2}{4\pi}\right)^{\epsilon} g^2C_F\delta(1-x) \nu^{\eta} \int \frac{d^{d}k\;\delta(k^2)}{(2\pi)^{d-1}} \frac{\delta(\tra{E}-\prp{k}) \Theta(\prp{k}-k^- r)}{k^+ (k^-)^{1+\eta}} \nn\\
  &= 2 \frac{\alpha_s C_F }{\pi}  \frac{(e^{\gamma_E} \mu^2)^\epsilon}{\Gamma(1-\epsilon)} \nu^{\eta} \delta(1-x) \frac{1}{\tra{E}^{1+2\epsilon}} \int_0^{\tra{E}/r} \frac{dk^-}{(k^{-})^{1+\eta}} \nn \\
  &= - 2 \frac{\alpha_s C_F }{\pi}  \frac{(e^{\gamma_E} \mu^2)^\epsilon}{\Gamma(1-\epsilon)} \frac{(\nu r)^{\eta}}{\eta} \delta(1-x) \frac{1}{\tra{E}^{1+2\epsilon+\eta}}.
\end{align}
Adding both contributions and expanding in $\eta$ keeping $\epsilon$ finite we have
\begin{align}
  \label{eq:totDBqq}
 & \Delta B_{q/q}^{(1)}(\tra{E},x,p^-,r)  =  \widetilde{\Delta B}_{q/q}^{(1)} -  \Delta B^{\text{ z-bin},(1)}_{q/q}  \\
  & = \frac{\alpha_s C_F }{\pi}  \frac{(e^{\gamma_E} \mu^2)^\epsilon}{\Gamma(1-\epsilon)}\frac{ (\nu r)^{\eta} }{\tra{E}^{1+2\epsilon+\eta}} \lbc \frac{2}{\eta}  \delta(1-x)  +   \lb \frac{1+x^2}{(1-x)^{1+\eta}} -\epsilon (1-x)^{1-\eta}  \rb  \left( \frac{\tra{E}}{p^- r} \right)^{\eta} \Theta(x-x_0) \rbc \nn \\
  & =\frac{\alpha_s C_F }{\pi}  \frac{(e^{\gamma_E} \mu^2)^\epsilon}{\Gamma(1-\epsilon)}  \frac{1}{\tra{E}^{1+2\epsilon}} \lbc (1+x^2) \lb \frac{\Theta(x-x_0)}{1-x} \rb_+ -2 \ln \Big{(} 1+ \frac{\tra{E}}{p^{-} r} \Big{)} \delta(1-x) -\epsilon (1-x) \rbc  \nn \;.
\end{align}
To get from the second line to the last line we used:
\begin{equation}
  \label{eq:plus1}
  \Theta(x-x_0)\mathcal{L}_0(1-x)=\lb \frac{\Theta(x-x_0)}{1-x} \rb_+ +\ln(1-x_0) \delta(1-x).
\end{equation}
A similar identity that will be used below is
\begin{equation}
  \label{eq:plus2}
  \Theta(x-x_0)\mathcal{L}_1(1-x)=\lb \frac{\Theta(x-x_0)}{1-x} \ln(1-x)\rb_+ +\frac{1}{2}\ln^2(1-x_0) \delta(1-x).
\end{equation}
The plus-functions on the right hand side of Eqs.(\ref{eq:plus1},\ref{eq:plus2}) are defined such that 
\bea
\int_0^1dx \lb \frac{\Theta(x-x_0)}{1-x} \rb_+ = \int_0^1dx \lb \frac{\Theta(x-x_0)}{1-x} \ln(1-x)\rb_+ =0. \nn
\eea

It should be noted that the last line of Eq.(\ref{eq:totDBqq}) does not contain any divergences in the simultaneous limit $\tra{E} \to 0$ and $x \to 1$, thus we can safely take the limit $\epsilon \to 0$. This gives the final result
\begin{equation}
 \Delta B_{q/q}^{(1)}(\tra{E},x,p^-,r)  =\frac{\alpha_s C_F }{\pi}  \frac{1}{\tra{E}} \lbc (1+x^2) \lb \frac{\Theta(x-x_0)}{1-x} \rb_+ +2 \ln (x_0) \delta(1-x)  \rbc + \mathcal{O}(\eta,\epsilon)\;.
\end{equation}

For the contribution of the gluon PDF to the quark beam function there is no soft-bin subtraction or rapidity divergences involved therefore we have:
\begin{align}
  \label{eq:calcDBqg}
   \Delta B_{q/g}^{(1)}(\tra{E},x,p^-,r)& = -\frac{1}{1-\epsilon}\frac{T_F}{C_F}\int d\Phi_2^{\text{c,ISR}}(t,x')\sigma_{2}^{\text{c}}(-t/x',(x'-1)/x') \delta(x-x') \Theta_{\text{meas.}} \nn \\
  &= \frac{\alpha_s T_F}{\pi (1-\epsilon)} \mu^{2\epsilon} \lb P_{qg}(x) -\epsilon \rb   \frac{1}{\tra{E}^{1+2\epsilon}} \Theta(x - x_0),
\end{align}
and since the final result does not contain any divergences in the simultaneous limit we have
\begin{equation}
   \Delta B_{q/g}^{(1)}(\tra{E},x,p^-,r)= \frac{\alpha_s T_F}{\pi }  \frac{1}{\tra{E}}  \lb P_{qg}(x) \rb  \Theta(x - x_0).
\end{equation}


\subsection{Jet-veto measurement}
Since at NLO there is only one parton contributing to ISR we can obtain the jet-veto measurement expressions by integrating the transverse energy results before performing the expansion in $\epsilon$ and $\eta$. In general the jet-veto measurements should depend on the jet radius $R^{\text{veto}}$ that appear in two-loop and higher order calculations. Since here we are considering only the one-loop contributions we will omit from the arguments this dependence on $R^{\text{veto}}$. From Eq.(\ref{eq:calcDBqq}) we have
\begin{align}
 \label{eq:calcDBqqV}
 \widetilde{\Delta B}_{q/q}^{(1)}(\veb{p},x,p^-,r) &=\frac{\alpha_s C_F}{\pi} \frac{(e^{\gamma_E} \mu^2)^\epsilon}{\Gamma(1-\epsilon)}\Big{(}\frac{\nu}{p^-} \Big{)}^{\eta}  \lb \frac{1+x^2}{(1-x)^{1+\eta}} -\epsilon (1-x)^{1-\eta}  \rb  \int_0^{\veb{p}} \frac{d\tra{E}\Theta(x-x_0)}{\tra{E}^{1+2\epsilon}} \nn \\
 &=\frac{\alpha_s C_F}{2\pi \epsilon} \frac{(e^{\gamma_E} \mu^2)^\epsilon}{\Gamma(1-\epsilon)}\Big{(}\frac{\nu}{p^-} \Big{)}^{\eta}  \lb \frac{1+x^2}{(1-x)^{1+\eta}} -\epsilon (1-x)^{1-\eta}  \rb \nn \\ &\times \Theta(x-x_0^{\text{vet.}})  \lbc \Big{(} \frac{x}{p^-r(1-x)} \Big{)}^{2\epsilon} - \frac{1}{(\veb{p})^{2\epsilon} } \rbc.
\end{align}
where $x_0^{\text{vet.}}=(1+\veb{p}/(p^- r))^{-1}$. Similarly for the soft-bin subtraction from Eq.(\ref{eq:calcDBqqS}) we have :
\begin{align}
  \label{eq:calcDBqqSV}
  \Delta B^{\text{ z-bin},(1)}_{q/q}(\veb{p},x,p^-,r) &= - 2 \frac{\alpha_s C_F }{\pi}  \frac{(e^{\gamma_E} \mu^2)^\epsilon}{\Gamma(1-\epsilon)} \frac{(\nu r)^{\eta}}{\eta} \delta(1-x) \int_0^{\veb{p}}\frac{d\tra{E}}{\tra{E}^{1+2\epsilon+\eta}} \nn \\
  &= 2 \frac{\alpha_s C_F }{\pi}  \frac{(e^{\gamma_E} \mu^2)^\epsilon}{\Gamma(1-\epsilon)} \frac{(\nu r)^{\eta}}{\eta} \delta(1-x) \lb \frac{1}{2\epsilon+\eta} \frac{1}{(\veb{p})^{2\epsilon +\eta}} \rb.
\end{align}
Adding both contributions and expanding first in $\eta$ and then in $\epsilon$ we get
\begin{multline}
  \Delta B_{q/q}^{(1)}(\veb{p},x,p^-,r) = \widetilde{\Delta B}_{q/q}^{(1)}(\veb{p},x,p^-,r) -\Delta B^{\text{ z-bin},(1)}_{q/q}(\veb{p},x,p^-,r) \\
   = \frac{\alpha_s C_F }{\pi} \lbc -\delta(1-x) \ln^2 \Big{(} \frac{\veb{p}}{p^- r} \Big{)} -(1+x^2)\lb \mathcal{L}_1(1-x)\\ - \mathcal{L}_0(1-x)  \ln \Big{(} \frac{x\veb{p}}{p^- r} \Big{)}  \rb \Theta(x-x_0^{\text{vet.}})\rbc.
\end{multline}
We can further simplify this result using Eqs.(\ref{eq:plus1}) and (\ref{eq:plus2})
\begin{multline}
  \Delta B_{q/q}^{(1)}(\veb{p},x,p^-,r) =\frac{\alpha_s C_F }{\pi} \lbc (1+x^2) \Big{(} \lb \frac{\Theta(x-x^{\text{vet.}}_0)}{1-x} \rb_+ \ln \Big{(}\frac{x\veb{p}}{p^- r} \Big{)} \\
  -\lb \frac{\Theta(x-x_0^{\text{vet.}})}{1-x} \ln(1-x)\rb_+ \Big{)}  -\ln^2(x_0^{\text{vet.}})\delta(1-x) \rbc.
\end{multline}
For the contribution of the gluon PDF to the quark beam function in jet-veto measurements we have by integrating Eq.(\ref{eq:calcDBqg}):
\begin{align}
  \Delta B_{q/g}^{(1)}(\veb{p},x,p^-,r)&= \frac{\alpha_s T_F}{\pi (1-\epsilon)} \mu^{2\epsilon} \lb P_{qg}(x) -\epsilon \rb  \int_0^{\veb{p}} \frac{d \tra{E}}{\tra{E}^{1+2\epsilon}} \Theta(x-x_0)  \\
  &=-\frac{\alpha_s T_F}{\pi} \lb \ln \left( \frac{p^{-}r}{\veb{p} x} \right)  + \ln(1-x)\rb P_{qg}(x) \Theta(x-x_0^{\text{vet.}}) + \mathcal{O}(\epsilon)\;. \nn
\end{align}




\section{Jet function contributions from out-of-jet radiation}
\label{app:C}

In this section we demonstrate how the contributions from out-of-jet radiation to the jet function in hadronic collisions and the total cross section at NLO are suppressed by a factor of $\tra{E}/p_T$ for transverse energy measurements and $\veb{p}/p_T$ for transverse veto measurements. The results we obtain are independent of the measurements within the jet-cone, thus our conclusions apply to both measured and unmeasured jets. We start the calculation with the expression of the jet function in Eq.(4.12) of Ref.~\cite{Ellis:2010rwa} inserting the appropriate transverse energy $\delta$-function,
\begin{multline}
  \label{eq:c1}
  \tilde {J}^{g\text{-out}}_{q}= g^2 \lp \frac{ e^{\gamma_{E}}\mu^2}{4 \pi}\rp^{\epsilon} C_F \int \frac{d\ell^+}{2 \pi} \frac{1}{(\ell^+)^2} \int \frac{ d^dq}{(2 \pi )^{d-2}} \lb 4 \frac{\ell^+}{q^-} +2 (1-\epsilon) \frac{\ell^+-q^+}{\omega-q^-}  \rb \delta(q^+ q^- -\tra{q}^{2}) \\
  \delta(\ell^+ -\frac{q^+ \omega}{\omega-q^-}) 
  \delta(\tra{E}-q_T) \Theta( q^+/q^- - (s_J R/2)^2 ),
\end{multline}
where $q_T \simeq  \vert \bmat{q} \vert \sin(\theta_J) \equiv  \vert \bmat{q} \vert s_{J}$ is the transverse momentum of the gluon escaping the jet with respect to the beam axis $\hat{n}_{B} =(c_J,s_J,0,...,0)$, with $c_i$ and $s_i$ the cosine and sine of the angle $\theta_i$ respectively. Since this gluon is a collinear in scaling, from power counting we have $\vert \bmat{q} \vert = q^-/2 + \mathcal{O} (\lambda^2 q^-)$. Performing the integrals $d\ell^+$ and $d^{d-2} \tra{q}$ using the first two $\delta$-functions in Eq.(\ref{eq:c1}) we get,
\begin{multline}
  \tilde {J}^{g\text{-out}}_{q}= 2\frac{\alpha_s C_F}{2 \pi} \frac{ (e^{\gamma_{E}}\mu^2)^{\epsilon}}{\Gamma(1-\epsilon)}  \int \frac{dq^+ dq^-}{(q^+ q^-)^{1+\epsilon}} \lb 1- \frac{q^-}{\omega}  +\frac{1}{2}(1-\epsilon) \frac{(q^-)^2}{\omega^2}  \rb  \\ \times \delta(\tra{E}-q_T) \Theta\lp q^+/q^- - (s_J R/2)^2\rp.
\end{multline}
Performing  the remaining integrals we have,
\begin{equation}
  \label{eq:c2}
   \tilde {J}^{g\text{-out}}_{q}= 2\frac{\alpha_s C_F}{2 \pi} \frac{ (e^{\gamma_{E}}\mu^2)^{\epsilon}}{\Gamma(1-\epsilon) \epsilon R^{2\epsilon}} \frac{1}{\tra{E}^{1+2\epsilon}} \lb 1- \frac{\tra{E}}{p_T} +\frac{1}{2} (1-\epsilon) \frac{\tra{E}^2}{p_T^2} \rb.
\end{equation}

Before we further  expand  Eq.(\ref{eq:c2}) we evaluate the zero-bin subtraction of the Eq.(\ref{eq:c1}) which is given by the following (see Eq(4.14) of Ref.~\cite{Ellis:2010rwa}),
\begin{multline}
  \label{eq:c3}
   J^{g\text{-out, z-bin}}_{q}= 4 g^2 \lp \frac{ e^{\gamma_{E}}\mu^2}{4 \pi}\rp^{\epsilon} C_F \int \frac{d\ell^+}{2 \pi} \frac{1}{(\ell^+)^2} \int \frac{ d^dq}{(2 \pi )^{d-2}}   \frac{\ell^+}{q^-}  \delta(q^+ q^- -\tra{q}^2) \delta(\ell^+ - q^+) \\ 
  \times \delta(\tra{E}-q_T) \Theta( q^+/q^- - (s_J R/2)^2 ).
\end{multline}
Following the same steps used to obtain Eq.(\ref{eq:c2}) we find,
\begin{equation}
  \label{eq:c4}
   J^{g\text{-out, z-bin}}_{q}= 2\frac{\alpha_s C_F}{2 \pi} \frac{ (e^{\gamma_{E}}\mu^2)^{\epsilon}}{\Gamma(1-\epsilon) \epsilon R^{2\epsilon}} \frac{1}{\tra{E}^{1+2\epsilon}} .
\end{equation}
We note that the zero-bin term exactly will cancel the first term in the square brackets of Eq.(\ref{eq:c2}) thus for our final result we have
\begin{equation}
   \label{eq:c5}
   \tilde {J}^{g\text{-out,b}}_{q}= -2\frac{\alpha_s C_F}{2 \pi} \frac{ (e^{\gamma_{E}}\mu^2)^{\epsilon}}{\Gamma(1-\epsilon)  R^{2\epsilon}} \frac{1}{\tra{E}^{1+2\epsilon}} \lb \frac{1}{\epsilon}\lp \frac{\tra{E}}{p_T} - \frac{\tra{E}^2}{2 p_T^2}  \rp + \frac{\tra{E}^2}{2p_T^2} \rb .
  \end{equation}
Expanding in $\epsilon$ and keeping only the leading order in $\tra{E}/p_T$ finite terms we have
\begin{equation}
  J^{g\text{-out,LP}}_{q}= 2\frac{\alpha_s C_F}{ \pi}  \frac{1}{\tra{E}} \lb \frac{\tra{E}}{p_T} \ln \lp\frac{\tra{E} R}{\mu} \rp  + \mathcal{O} \lp \frac{\tra{E}^2}{p_T^2} \rp\rb.
\end{equation}
Since the jet canonical scale is $\mu_{J} = \tra{p} R$, for $\tra{E} \ll \tra{p}$ this term is suppressed compared to the leading contributions of the corresponding $n_J$-collinear soft function in Eq.(\ref{eq:CSnJ}) and therefore maybe ignored during the computation of jet cross section. Similar suppression is found for the integrated (jet-veto) measurement.


\bibliography{paper}

\end{document}